\documentclass[preprint]{aastex}	 
\usepackage{epsfig,emulateapj5}

\newcommand{\km}{${\rm km\,s}^{-1}$}
\newcommand{\fuse}{\em FUSE\/}

\newcommand{\vinf}{${v}_\infty$}

\slugcomment{Version \today}
\shorttitle{Survey of \ion{O}{6} wind variability in OB-type stars}
\shortauthors{Lehner et al.}
\begin{document}

\title{{\em Far Ultraviolet Spectroscopic Explorer}\/
Snap-Shot Survey of \ion{O}{6}  Variability in the Winds of 66 OB-Type Stars}

\author{N.\ Lehner,\altaffilmark{1,2}
   A.\ W.\ Fullerton,\altaffilmark{1,3}
   D.\ Massa,\altaffilmark{4}
   K.\ R.\ Sembach,\altaffilmark{5} and
   J.\ Zsarg\'o\altaffilmark{1}
   }

\altaffiltext{1}{Department of Physics and Astronomy, 
                 The Johns Hopkins University,
                 3400 N. Charles Street, 
		 Baltimore, MD 21218.
		 }

\altaffiltext{2}{Present address: Department of Astronomy, University
of Wisconsin, 475 North Charter Street, Madison, WI 53706. nl@astro.wisc.edu}

\altaffiltext{3}{Department of Physics and Astronomy, 
                 University of Victoria, 
		 P.O. Box 3055,
                 Victoria, BC V8W 3P6, 
		 Canada.}

\altaffiltext{4}{SGT Inc.,
                 NASA's Goddard Space Flight Center, 
		 Code 681, Greenbelt, MD 20771.}

\altaffiltext{5}{Space Telescope Science Institute,
		 3700, San Martin Drive, Baltimore, MD 21218}

\shortauthors{Lehner et al.}
\shorttitle{\ion{O}{6} Wind in OB-Type Stars}

\begin{abstract}
We have used the {\em Far Ultraviolet Spectroscopic Explorer (FUSE)}\/ to conduct a 
snap-shot survey of \ion{O}{6}  variability in the winds of 66 OB-type stars
in the Galaxy and the Magellanic Clouds.
These time series consist of 
two or three observations separated by intervals ranging from  a few days to several months.
Although these time series provide the bare minimum of information
required to detect variations,
this survey demonstrates that the \ion{O}{6} doublet in the winds of OB-type stars
is variable on various scales both in time 
and velocity.  For spectral types from O3 to B1, 64\%
vary in time. At spectral types later than B1, no wind 
variability is observed. 
In view of the limitations of this survey, this fraction represents a
lower limit on the true incidence of variability in the \ion{O}{6} wind lines,
which is very common and probably ubiquitous.
In contrast, for \ion{S}{4} and \ion{P}{5}, only 
a small percentage of the whole sample shows
wind variations, although this may be principally due to selection effects.
The observed variations extend over several hundreds of 
\km\ of the wind profile and can be strong.
The width over which the wind \ion{O}{6} profile varies is only weakly
correlated with the terminal velocity ({\vinf}), but a significant correlation 
(close to a 1:1 relationship) is derived between the maximum
velocity of the variation and {\vinf}. High velocity \ion{O}{6} wind absorption features
(possibly related to the discrete absorption components seen in other wind lines)
are also observed in 46\% of the cases for spectral types from O3 to B0.5. 
These features are variable, but the nature of their propagation cannot
be determined from this survey.
If X-rays can produce sufficient \ion{O}{6} by Auger ionization of \ion{O}{4}, and the
X-rays originate from strong shocks in the wind, this 
study suggests
that stronger shocks occur more frequently near {\vinf}, 
causing an enhancement of \ion{O}{6} near {\vinf}. 
\end{abstract}


\section{Introduction}
The detection of wind profiles in the {\ion{O}{6}} 
$\lambda\lambda$1032, 1038 resonance doublet in {\it Copernicus} spectra of 
stars with spectral types between O4 and B1 {\citep{snow76,morton79}} provided
the first evidence for the existence of high-energy, non-radiative processes 
in the outflows from hot stars.
The presence of this ion was surprising, since only the hottest O 
stars are expected to produce it directly through photoionization.
In a pivotal paper, {\citet{cassinelli79}} demonstrated that X-rays can
produce sufficient {\ion{O}{6}} by Auger ionization of {\ion{O}{4}}, 
which is the dominant ionization stage of oxygen in the winds of stars
in the temperature range where {\ion{O}{6}} is observed.
The X-ray flux required by the Auger mechanism was subsequently detected in 
spectra of OB-type stars obtained by the {\it Einstein} observatory 
{\citep{harnden79,seward79}}.
Since the existence of {\ion{O}{6}} is linked directly to the presence
of X-rays, the shape and strength of the {\ion{O}{6}} lines can be
used to trace the distribution of hot, X-ray emitting gas in 
the winds of early-type stars {\citep{macfarlane93}}.

X-rays are believed to be caused by the formation of strong shocks in 
the wind, which convert mechanical energy of the flow into localized 
sources of heat.
These shocks could arise from various processes, such as:
(a) the ``line de-shadowing instability'' intrinsic to the 
    line-driving mechanism responsible for hot-star winds
    {\citep{owocki88}};
(b) the interfaces of large-scale co-rotating interactions regions
    (CIRs), which might be responsible for recurrent variability in
     hot-star winds {\citep{mullan84,cranmer96}}; or
(c) collisions (channeling) of material along closed (open) magnetic
    fields emanating from the surface of the star
    {\citep{uddoula02}}.
Since these mechanisms involve the evolution of complicated, 
non-stationary flows, their tracers might be expected to 
exhibit substantial variations.
For example, if the bulk of the X-rays or {\ion{O}{6}} ions are
produced by a few strong shocks (as in the case of CIRs and simple
magnetic field configurations), large variations might be expected as the 
distribution or strength of the shocks evolves.
Conversely, little or no variability would be expected if the
global distribution of the shocks (e.g., resulting from many
ensembles of shocks generated by the line-deshadowing instability) do not
change much with time.
In either case, fluctuations in the tracers of the high energy
processes provide information about the origin and distribution 
of the hottest gas in these outflows.

Until recently, X-rays were the predominant tracers of shock
phenomena in the winds of hot stars.
The limited X-ray data available suggest little short-term variability,
which is generally interpreted as implying that many shock ensembles
are present in the wind {\citep[see,~e.g.,][]{feldmeier97}}, so
that the evolution of any one of them is of little consequence.
However, a variety of selection effects might also bias this
interpretation; see, e.g., {\citet{oskinova01}}.
At a fundamental level, spatially localized variations in the X-ray flux
may be difficult to detect against the background contributions collected
from nearly the entire volume of the wind.

This situation changed dramatically with the launch of the 
{\it Far Ultraviolet Spectroscopic Explorer (FUSE)\/} satellite,
which provides routine access to the resonance lines of 
{\ion{O}{6}}.
Observations of these wind features permit more sensitive
searches for variability, primarily because they are formed by the
resonant scattering of photons in localized regions of the wind 
(i.e., the column of material projected against the disk of the star).  
Such observations also permit wind features to be identified, even though
they are otherwise difficult to detect because of blends with 
strong stellar and interstellar lines. 
Moreover, the {\it FUSE\/} spectrographs are themselves very sensitive,
so that OB-type stars in the Magellanic Clouds are easily accessible.

In order to exploit this new capability, we designed several 
Principal Investigator (PI) Team programs to obtain sparse time series
observations for a large sample of OB-type stars in the Milky Way and
Magellanic Clouds.
These time series were obtained by breaking long integrations into
two or three observations separated by intervals that
typically amounted to a few days.
Although these time series provide the bare minimum of information
required to detect variations, they permitted a broad range of
stellar parameters to be surveyed.
The aim was to provide a rudimentary assessment of the frequency
of {\ion{O}{6}} variability as a function of spectral type
and luminosity class; to determine whether {\ion{O}{6}} was
as variable as the stellar wind features in \ion{P}{5} and \ion{S}{4};
and to glean whatever information possible about the nature of the variations.
Although this approach is statistical in nature, it complements other 
studies undertaken with {\it FUSE\/} designed to characterize stellar wind 
variability in {\ion{O}{6}} and other far ultraviolet wind lines for specific
targets {\citep[e.g.,][]{fullerton03}}.
It also provides the information required to assess the effect
of stellar wind variability on measurements of the interstellar
{\ion{O}{6}} lines {\citep[see,~e.g.,][]{lehner01}}. 

\section{{\fuse} Observations}
\subsection{The sample}
In Tables~\ref{t1} and \ref{t2}, we summarize the Milky Way (MW) 
and Magellanic Clouds (MC) observations, respectively. Basic properties (Galactic 
coordinates,  spectral-type, magnitude and reddening, 
terminal\footnote{The terminal velocity of a stellar wind is defined as 
the velocity of outflowing matter at large distances from the star,
where it is no longer experiencing significant acceleration but 
is not yet interacting with the interstellar medium \citep[e.g.,][]{prinja90}.}
and projected rotational velocities) are indicated. Columns 7 and 8 
show the separation in days between two successive observations and 
the dates when they were taken, respectively. Most of 
these sparse time series  consist of 2--3 observations separated 
by intervals ranging from a few days to several months.
The ninth column identifies the rootname of the {\fuse} data set,
and is followed by remarks concerning the presence of radial
velocity variations in the object.

The time-series
observations were obtained mainly through two large
programs conducted by the {\fuse}\/ PI Team, namely the 
``\ion{O}{6} ISM" (P101, P102, P122) and ``hot stars" (P117) working group investigations. 
The purposes of these multiple observations were to investigate
the \ion{O}{6} time-variability in the winds of early-type stars
from a stellar point of view (this paper) and to assess its impact on the
interstellar absorption measurements of \ion{O}{6} \citep{lehner01}.
Other (non-proprietary) data from Guest Investigator programs were also 
included (observation rootnames starting with A and B), as were targets 
obtained as early-release observations  (rootnames starting with X). 

Table~\ref{t1} indicates that the MW sample
is strongly biased toward later type stars (O9-B0-B0.5), while the
MC sample (Table~\ref{t2}) is slightly more uniformly distributed but 
of smaller size. 
These selection effects limit the conclusions that can be drawn: the sample is
biased in spectral type; it is not large enough for a meaningful statistical
analysis to be performed for each
spectral and luminosity class; and the number of multiple 
observations is very small. It is nonetheless the largest
sample available  to study \ion{O}{6} variability in the winds of early-type stars. 
A total of 66 stars (44 Galactic stars, 20 LMC stars, and
2 SMC stars) were observed with typically 2 or 3 
observations, except for a few cases with 4 exposures.
A future complementary study  will involve the analysis of 
the high time-resolution observations of two LMC stars \citep{fullerton03}.

\subsection{{\fuse} Instrument and Data Reduction}
{\fuse} consists of four co-aligned telescopes and spectrographs, termed ``channels'', two of which have 
SiC coatings to maximize the sensitivity in the wavelength range $\sim$905--1105 \AA,
while the other two have Al-LiF coatings for sensitivity in the 
$\sim$1000--1187 \AA\ range. 
This wavelength region provides access to a rich suite of resonance
and excited lines, in particular for the ions considered here: \ion{O}{6}, \ion{S}{4}, and \ion{P}{5}.
The channels are referred
to as SiC\,1 and SiC\,2 and LiF\,1 and LiF\,2, respectively.
This design includes overlap at certain wavelengths, which helps to
distinguish between real features in the spectra and instrumental
artifacts, particularly fixed-pattern noise. Since we are interested in 
time-variability in the spectra, this redundant coverage is essential 
to make sure the differences are in the spectrum of the source and not due to 
detector defects. 
More complete descriptions of the 
design and performance of the {\fuse}\/ spectrographs are given 
by  \citet{moos00} and \citet{sahnow00}, respectively.

Most of the  observations were obtained through the $30\arcsec \times 30\arcsec$
(LWRS) apertures, except for a few cases where they were obtained 
through the $4\arcsec \times 20\arcsec$ (MDRS) and $1\farcs 25 
\times 20\arcsec$ (HIRS) apertures. Since flux can be lost through the
narrower apertures, systematic multiplicative corrections were 
applied to adjust the flux levels to a common value. These 
corrections do not affect the detection of variations.
In a few cases, however, and especially
for LiF\,1B or LiF\,2A (which contain the \ion{P}{5} lines), the adjustments
are compromised by contamination from the ``worm,'' 
a feature that deforms the continuum non-linearly. The worm is understood 
to be due to shadowing from grid wires in front of the detector \citep{sahnow02}.

Standard processing with  version 2.0.5 of the CALFUSE calibration
pipeline software was used to extract
and calibrate the spectra. The software screened the data
for valid photon events, removed burst events, corrected
for geometrical distortions, spectral motions, satellite orbital
motions, and detector background, and finally applied
flux and wavelength calibrations.
The extracted spectra associated with the separate exposures of a given
observation were aligned by cross-correlating the positions of
strong interstellar lines, co-added, and rebinned to a spectral resolving power of
$\sim 15,000$ ($\sim 20$ \km).

\subsection{The Ions Under Study: \ion{O}{6}, \ion{S}{4}, and \ion{P}{5}}\label{ions}
To compare to wind variability in \ion{O}{6}, we chose to examine the lines of 
\ion{S}{4} and \ion{P}{5}
based on two main criteria: (i) They are present over a wide range of spectral types (i.e., 
\ion{S}{4} and \ion{P}{5} are dominant ionization stages or near 
dominant ionization stages, see below). (ii) The lines
fall in sections of the spectrum where they are minimally contaminated by
blending with interstellar features. At
wavelengths below 1000 \AA, the signal-to-noise level becomes 
significantly weaker because the effective areas of the SiC channels are smaller. 

Below, we briefly recall some properties of 
the ions under consideration and their behavior through
the H-R diagram of OB-type stars. This is mostly based on the two recent {\fuse}\/
atlases of OB-type stars in the MW \citep{pellerin02}
and the MC \citep{walborn02}
 
Owing to the high ionization required to form \ion{O}{6} (I.P.: 113.9--138.1 eV),
the \ion{O}{6} $\lambda$$\lambda$1031.926, 1037.617 resonance doublet is the most useful tracer
of high-energy processes in the optical/UV region of the spectrum.
Its prevalence is believed
to be directly related to the distribution of X-rays in the 
winds of the hot stars. In this sense, \ion{O}{6} is often referred to as 
a ``super-ion'' \citep[see, e.g.,][]{massa02}. Other super-ions exist in FUV spectra
(e.g., \ion{N}{5} and \ion{S}{6}), but \ion{O}{6} remains unique because
it is two stages above the dominant ionization stage of \ion{O}{4}.
Because of this, Auger ionization of \ion{O}{4} is believed
to be the dominant process for producing \ion{O}{6} \citep{cassinelli79}.

Throughout the early O-type dwarf stars and from O2 to B0 supergiants, 
the \ion{O}{6} lines exhibit typical P Cygni profiles \citep{pellerin02}, which consist of an extended 
blue-shifted absorption trough and red-shifted emission peak. 
Because the separation between the \ion{O}{6} lines is only 1650 \km,
the absorption trough of \ion{O}{6} $\lambda$1038 is often blended 
with the emission part of \ion{O}{6} $\lambda$1032. Similarly, the absorption trough of
\ion{O}{6} $\lambda$1032 can be lost in the saturated absorption line of
\ion{H}{1} Ly\,$\beta$. 
Moreover, several other interstellar absorption
lines (\ion{C}{2}, \ion{O}{6}, and H$_2$) can complicate the appearance 
of stellar \ion{O}{6} line profiles. We discuss the different contaminations of the wind
profiles in more detail in Section~\ref{tv}.
One should refer to the normalized spectra in the MC {\fuse}\/ atlas of OB-type stars
\citep{walborn02}, and to the unnormalized spectra 
of the MW {\fuse}\/ atlas of OB-type stars \citep{pellerin02} for 
illustrations of these spectra as a function of spectral type.
%
%

\ion{P}{5} $\lambda$$\lambda$1117.977, 1128.008 
(I.P.: 51.4--65.0 eV, similar to \ion{He}{2}) is a resonance
doublet expected to be near the dominant ionization
stage in the winds of O-type stars \citep{massa02}. It weakens substantially 
in spectra of B0 and later type stars, for which \ion{P}{4} becomes the dominant
ion.  Even though it is a dominant
ion, it remains unsaturated because of its very low cosmic abundance
($12 + \log[{\rm P/H}] = 5.57$  compared to 8.87 for O and and 7.27
for S). This doublet has a luminosity dependence. In mid-O spectra,
it appears mainly as photospheric absorption in dwarfs, a weak P Cygni
profiles in giants, and strong P Cygni profiles in supergiants
\citep{pellerin02}. 

\ion{S}{4} $\lambda$$\lambda$1062.664, 1072.973 
(I.P.: 34.8--47.3 eV) is one stage below 
the expected dominant ion stage of S in the winds of early O-type stars, but
starts to dominate in late O and early B-type stars. \ion{S}{4} 
$\lambda$1063 is a resonance line, while \ion{S}{4} $\lambda$1073
arises from low-lying excited levels that produce two transitions at wavelengths 
1072.973 and 1073.516 \AA. Since the transition at 1072.973 \AA\
is about 9 times stronger than the transition at 1073.516 \AA,
we consider only the transition at 1072.973 \AA. 
These transitions have a similar luminosity effect to 
the \ion{P}{5} doublet for mid-O spectral types.  However, in the
spectra of early B-type stars, the \ion{S}{4} lines are nearly independent 
of luminosity \citep{pellerin02}.

We note that even though the ion fraction of both \ion{P}{5} and 
\ion{S}{4} could be substantially larger than the ion fraction
of \ion{O}{6}, their low abundance (0.05\% and 2.5\% as abundant 
as O, respectively) could make their detection difficult, especially in the MC 
where the metallicity is down by about $\ga 50$\% compared to the MW. 

\subsection{Detecting Time-Variability with {\fuse}}\label{tv}

Several careful steps were taken to uncover time-dependent
changes in the profiles of these sparse time series 
observations:
\\
(i) For each species, we first transformed the spectra from wavelength space 
to velocity space in order to facilitate direct comparisons between
the components of the doublet and lines from different species.
This velocity is in the heliocentric frame,
but, unfortunately, the  velocity zero point in 
{\fuse}\/ spectra remains uncertain by several tens of \km.
The relative velocity can be accurate to a few \km\  within
one channel, but can vary by more than 10 \km\ between channels.
Using the velocity scale, 
we compared both the calibrated fluxes and the ratio of the fluxes.
\\
(ii) The blue component of the \ion{O}{6} doublet is affected by
the airglow emission and the interstellar absorption of \ion{H}{1} Ly\,$\beta$,
which lies 1805 \km\ blueward of {\ion{O}{6}} $\lambda$1032. In 
Tables~\ref{t1} and \ref{t2}, we give the terminal velocity,
$v_\infty$, which should give roughly the extent of the wind 
profiles.\footnote{Positions in the absorption trough of a P~Cygni profile occur at
negative (blueshifted) velocities.  To simplify the notation, we
have expressed them as positive quantities; i.e., multiplication
by $-1$ is implied.}
When $v_\infty$ is greater than 1805 \km, the blue 
component of the \ion{O}{6} doublet is completely absorbed by
the interstellar \ion{H}{1} Ly\,$\beta$ line.
Therefore, the  \ion{O}{6} $\lambda$1038 component was first checked for possible
variations. If the ratio of the fluxes in the time series 
is a horizontal line, we conclude that the wind profiles
were the same at the times when the observations occurred.
Note that
for the red component of the doublet
there is generally no information at velocities below 400--500 \km\ because of 
blending with the saturated interstellar
\ion{C}{2} $\lambda$1036 line.
\\
(iii) If there were significant changes in the wind profiles between the different observations, we 
cross-checked that the profiles (and the ratio of the fluxes) were 
the same in the redundant spectra extracted from different channels of the observations. 
The \ion{O}{6} $\lambda$$\lambda$1032, 1038 resonance doublet 
appears in LiF\,1A, LiF\,2B, SiC\,1A, and SiC\,2B spectra. Because the 
mirrors and gratings move slightly and can be misaligned, 
information in one channel 
is occasionally lost, but usually there is more than one channel available. 
This step is especially important with {\fuse}\/ observations because the
aging of detector components can introduce shifts 
and distortions in the spectral features (known as the ``walk") that could mimic
the variations we are looking for. However,
this effect is larger near the edge of detector segments SiC\,1A and SiC\,2B and
rather small near the center of detector segments LiF\,1A and LiF\,2B.
The walk problem is also not expected to occur at the same position
with the same amplitude in the different detector segments. Thus, by carefully comparing the
wind profiles from different segments, we can determine if the observed variation 
is intrinsic to the stellar wind. 
\\
(iv) Further information comes from the \ion{O}{6} $\lambda$1032 transition
when this line is not too contaminated by the \ion{H}{1} Ly\,$\beta$
airglow or absorption  or \ion{O}{1} airglow.  We required 
that when \ion{O}{6} $\lambda$1032 is available, it exhibits
similar variations at the same velocities as the $\lambda$1038 component.
If the wind profiles are unsaturated, we can also check that
the line strength of \ion{O}{6} $\lambda$1032 is approximately twice the value 
of \ion{O}{6} $\lambda$1038, as expected for optically thin absorption. 
Since the velocity separation between the \ion{O}{6} doublet is 
only 1650 \km, there are cases where the wind profiles can overlap.
\\
(v) Similar procedures applied for \ion{S}{4} and  \ion{P}{5}, but because
the doublet separations of these species are much larger than
for \ion{O}{6} (2900 and 2480 \km, respectively), there is generally
no problem of overlapping wind profiles. 
There is also less blending with  saturated interstellar lines. 
When strong H$_2$ lines are present, they can contaminate
\ion{S}{4} $\lambda$1073 at velocities greater than 2000 \km\ and 
\ion{S}{4} $\lambda$1063 at velocities less than 200-500 \km.
For \ion{P}{5}, there are generally no saturated interstellar lines,
but possible blends with photospheric \ion{O}{4} $\lambda$$\lambda$1122.3, 1124.9 
and \ion{Si}{4} $\lambda$$\lambda$1122.5, 1128.3  can complicate
the interpretation of these profiles. We note that
the variations of \ion{S}{4} $\lambda$1073 generally appear
larger than for \ion{S}{4} $\lambda$1063, contrary to the 
expectation based on their rather similar oscillator strengths.
A similar problem with the relative oscillator strengths of these
transitions was noted  by \citet{massa02}.

These procedures ensured that the observed variations 
occur in the stellar wind profile, and also that weak
variations could be detected in spectra characterized by signal-to-noise ratios of $\sim 20$.
Figure~\ref{fig1} shows 3 examples from the Galactic sample, for which we overplotted 
the spectra taken at two or three different times. The differences of 
these fluxes are also plotted
for \ion{O}{6} $\lambda$$\lambda$1032, 1038, \ion{S}{4} $\lambda$$\lambda$1062, 1073,
and \ion{P}{5} $\lambda$$\lambda$1118, 1128 versus velocity. 
In the following figures, we instead plot $\Delta \tau \tbond \ln[F(t_1)/F(t_2)]$ as a function of $v$,
which has  more physical meaning than a flux difference because it is related to
the optical depth of a wind absorption . 

The left-hand panel of Figure~\ref{fig1} shows an example where no variation is 
detected. We note a slight change in the difference of the profiles  
around $-3100$ \km, probably near the terminal velocity of this star. However,
no such change is observed in spectra from other segments, and we therefore rule out 
wind variability in this case. The right-hand panel
shows a large amplitude variation that extends over a large fraction of all profiles.
The middle panel shows a very small variation that extends
over $\sim$500 \km\ only in the \ion{O}{6} profile. In this case the redundant information
provided by both components of the doublet permits small variations in comparatively noisy 
spectra to be detected unambiguously. Note that in both the middle and right-hand panels,
the effect of the worm can be observed in \ion{P}{5} $\lambda1128$ for velocities $> -1000$ \km.
Yet by following the above procedure, we can distinguish between wind variability and 
instrumental artifacts.  See also \citet{lehner01} for other
examples. 

\subsection{Radial Velocity Variations}\label{binary}

Wholesale shifts in the positions of photospheric lines indicate
systematic motion of the star, possibly in a binary system.
Such motion produces a characteristic ``S'' shape in ratios or
differences of spectra at a location that depends on the shift in
radial velocity between the two spectra, and over a wavelength
region that depends on the width of the photospheric line.
We typically noted variations of this sort near the systemic
velocity of the star, which indicates only small radial velocity variations.
Figure~\ref{fig2} (upper-left panel) shows an example of such an ``S''-shape
near the systemic velocity (0 \km) in the profiles of {\ion{S}{4}} and
{\ion{P}{5}}, which are largely photospheric.
These variations are not seen in the {\ion{O}{6}} line, which instead
shows substantial variations between $-1550$ and $-750$~\km;
i.e., in the outflowing wind.
An entry of ``RV var'' (RV var?) in the last column of Tables~\ref{t1} and \ref{t2} indicates
the presence (suspected presence) of photospheric radial velocity variations
in the {\fuse}\ time series.

\section{Time Dependence in the \ion{O}{6} Wind Profiles}\label{td}
In Figure~\ref{fig2},  the logarithm of the ratio of the 
fluxes observed at different times is presented as a function of the velocity
for 4 examples in the MW and in the LMC, 
for \ion{O}{6} $\lambda\lambda$1032, 1038, \ion{S}{4} $\lambda\lambda$1062, 1073, and 
\ion{P}{5} $\lambda\lambda$1118, 1128. Based on analogous figures constructed for the whole sample, 
we summarize in Table~\ref{t3} our assessment of whether
the wind profiles for these ions were variable
at the time of these observations. We also indicate the velocity range over which 
the variation is observed. In some cases, the high velocity value for \ion{O}{6}
was derived using the other species as there is overlap of the \ion{O}{6} wind profiles 
at velocities larger than 1650 \km. For some cases -- particularly in the MC sample -- no information is
given concerning the velocity extent of the \ion{O}{6} wind variations 
due to uncertainties resulting from the lower signal-to-noise of the data and the greater degree of
contamination by \ion{H}{1} and \ion{O}{1} airglow emission lines.
The ``(n)" in Table~\ref{t3} indicates that the spectra suffer either from low
signal-to-noise level or some problem in the fluxes,  
or that the wind is contaminated by other stellar lines (in the 
case of radial velocity variables, and particularly for the \ion{P}{5} lines). Thus 
when ``(n)" is listed, detections of variability are more uncertain.

Several inferences can be drawn from these measurements: 

(1) For the whole sample, 56\% of the \ion{O}{6} wind profiles vary with time.
For the MW $24/44$ (55\%) and $13/22$ (59\%) for the MC 
show variability.  The frequency of \ion{O}{6} wind variability is 
therefore similar in both samples.
If we remove 8 stars with spectral 
type later than B1 for which \ion{O}{6} is unlikely to be present
\citep{morton79,cassinelli79,zsargo02}, 64\% of the combined sample is variable. 
In contrast, only 15\% and 5\% of the whole
sample vary in the wind profiles of \ion{S}{4} and \ion{P}{5}. These latter
percentages should be treated cautiously because the strengths of these ions vary
with spectral type and luminosity class (see \S~\ref{ions}). They are
expected to be strong in the winds of supergiants, and for those stars, 
the profiles are often variable (see Tables~\ref{t1} and \ref{t2}). 

(2) Figure~\ref{fig3} shows the fraction of stars for which we 
observed \ion{O}{6} wind changes as a function of spectral types and 
luminosity classes, where the error bars are $1/\sqrt N$, with $N$ the
total number of stars in a given bin. The sample is not large enough to draw any definite
conclusions, but suggests that the incidence of \ion{O}{6} variability increases for 
earlier spectral types. This, however, could also be a selection effect
caused by the stronger appearance of the P Cygni profiles in the  earlier-type stars. 

When all the spectral types are considered, all luminosity classes have a
similar percentage of \ion{O}{6} wind variability. 
However, if we take a more homogeneous sample and 
consider only spectral types between O9 and B0.5, 
we find that the \ion{O}{6} wind profiles vary in 7/10 (70\%) 
cases for supergiant stars (luminosity class I), 
8/12 (67\%) cases for giant stars (luminosity class II-III), 
and 6/13 (46\%) cases for the dwarf
stars (luminosity class IV-V). The results are, however, 
not statistically significant at the $1 \sigma$ level, and are insufficient to claim 
that \ion{O}{6} wind variability occurs
more frequently in supergiant and giant stars compared to dwarfs.
We note that for \ion{S}{4} and \ion{P}{5}, variability is observed only in 
supergiants (except one case observed in a class II star). 
This is not surprising because in less 
luminous stars these lines are predominantly photospheric.

(3) From Table~\ref{t3}, the full-width of the variation in the \ion{O}{6} profiles ranges from
approximately 225 \km\  to 2200 \km, but most fluctuations occur between $\sim$400 \km\ and 
1100 \km. The sample is not large enough to study spectral type
or luminosity dependence; but in any case no dependence is indicated.

(4) The strength and velocity of the variation can change over times as short as $\sim 1$
day. However, detailed observations with high-time resolution are required to
characterize the time scales rigorously.

(5) In Figure~\ref{fig4}, the full-width of the observed variations 
($\Delta v$), and the maximum absolute radial velocity 
where significant variability occurs ($v_{\rm max}$) are 
plotted as a function of {\vinf}. To investigate possible correlations,
we used the Spearman rank-order
correlation test, which has the advantage that the statistical 
significance of a non-zero rank correlation can be quantified reliably
without any assumption concerning the distribution of uncertainties.
We denote the rank-order correlation coefficient by $r$, and its
statistical significance by $t$.  The statistical significance is
defined in the interval [0,1], where small values indicate greater
significance.
\\
Although the top panel of Figure~\ref{fig4} suggests some correlation between $\Delta v$ and {\vinf}, 
the Spearman's  correlation test gives
$r = 0.73$  with  $t = 0.016$ (for the MW sample;   $r =  0.46$ with $t =  0.085$ for the whole sample). 
The width over which the wind profile of the \ion{O}{6} varies is only weakly
correlated with the terminal velocity. 
A least-squares fit to the MW sample gives 
$\Delta v  =(0.42 \pm 0.13)\, v_\infty + (170 \pm 245)$ \km\
with a minimum $\chi^2 = 13.7$ and a goodness of fit of 0.09. 
\\
In contrast, there is a significant correlation between $v_{\rm max}$
and {\vinf} as  $r = 0.92$ with  $t = 1.6 \times 10^{-4}$
(for the MW sample;   $r = 0.85$ with  $t = 6.4 \times 10^{-5}$ for the whole sample).  
A least-squares
fit to the data (MW and MC) gives $v_{\rm max}  =(0.90 \pm 0.11)\, v_\infty + (152 \pm 140)$ \km\
with a minimum $\chi^2 = 9.5$ and a goodness of fit of 0.31. 
The trend is close to the 1:1 relationship as illustrated 
in Figure~\ref{fig4}.

(6) The absolute optical depth variation can be characterized approximately by defining
$\Delta T \tbond |\Delta \tau_{\rm high} + \Delta\tau_{\rm low}|$, 
where $\Delta \tau_{\rm high}$ is the maximum positive 
logarithmic ratio of the fluxes observed at a specific velocity, and
$\Delta \tau_{\rm low}$ is the maximum absolute negative 
logarithmic ratio of the fluxes associated with the variation.
We find  a wide range of values for \ion{O}{6} $\lambda$1032: $0.4 \la \Delta T \la 1.0 $. 
The detection of $\Delta T \la 0.1$--0.2 is precluded by the noise level in these data. 

(7) $\Delta \tau$  does not seem to 
depend on the luminosity class or spectral type of the star
(see Figure~\ref{fig2}). The 
largest values of $\Delta \tau$ are observed for stars in the LMC.  (Note that
the $y$-scale in Figure~\ref{fig2} is larger for the LMC compared to 
the MW.)

\section{\ion{O}{6} Wind Absorption Features}\label{dac}
In the previous section, we assessed the frequency of variability
in the \ion{O}{6} wind line for many OB-type stars.
Extensive studies
of the winds through ultraviolet resonance lines in other species 
show that narrow or discrete absorption components (DACs) are associated 
with the extended absorption troughs \citep[for a recent review, see][]{prinja98a}.
These DACs become particularly strong near {\vinf}.
The typical P Cygni profiles observed for lower ionization stages
(e.g., \ion{C}{4}, \ion{Si}{4})
are also observed for the \ion{O}{6} doublet \citep{pellerin02,walborn02}.
Figure~\ref{fig5} shows that strong, high velocity absorption components 
are also observed in \ion{O}{6}. 

Several pieces of evidence confirm that these absorptions occur in 
the wind: 
(i) no strong photospheric lines were identified in this region of the spectrum
(we note, however, potential photospheric \ion{Fe}{3} and \ion{P}{3} 
lines that can contaminate the \ion{O}{6} spectrum for the later type stars;
see Zsarg\'o et al. 2003);
(ii) the velocity of the features changes from one observation to the next
\citep[for the best case see Figure~3 in][]{lehner01}; 
(iii) both components of the \ion{O}{6} doublet can be observed (see Figure~\ref{fig5}
and Lehner et al. 2001); 
and (iv) in cases of a binary, aligning the photospheric
lines do not align these features. The projected rotational velocity
also broadens the photospheric lines, and there is no correlation 
between the width of these absorption features and the $v \sin i $ values listed in Table~\ref{t1}.
Finally, these high-velocity features do not have 
an interstellar origin because in some cases 
they are too broad. Furthermore, transient features are not 
commonly observed in the interstellar gas. These features are therefore
due to \ion{O}{6} absorption in the stellar wind.

In Table~\ref{t4}, we present the stars for which we observed unambiguously
the \ion{O}{6} wind absorption features and their blue-shifted 
velocities, $v_{\rm abs}$, which correspond
to the deepest part of these 
features.\footnote{Not all the stars with absorption features 
exhibited variability.  For these cases, we used the morphological
approach described by \citet{zsargo02} to determine whether the
absorption features were formed in the wind.  For a positive detection,
we required that DACs be present at the same velocity as the
\ion{O}{6} components in archival {\it IUE} or {\it HST} spectra of the wind profiles
of one or more of the resonance lines of \ion{C}{4}, \ion{Si}{4}, and \ion{N}{5}.}
(When the feature has a flat bottom, its
centroid is used). The last column also presents the strength 
of these features, classified as weak, medium and strong. A ``weak" (small width and
small optical depth) absorption
corresponds to features as observed in the spectra of HD\,161807;
a ``medium" (intermediate) absorption corresponds to features as 
observed in the spectra of CPD$-$72\degr1184; and
a ``strong" (large width and large optical depth) absorption corresponds to features as observed in the spectra of HD\,92554
(see Figure~\ref{fig5}).
Sometimes weak and strong features are observed in the same spectrum. This is the 
case, e.g., for HD\,168941, for which the \ion{O}{6} $\lambda1038$ wind feature is blended with 
the \ion{O}{6} $\lambda1032$ absorption interstellar line. We take this qualitative approach rather than a quantitative
approach because the continuum is very uncertain for these generally complex
features (but see Zsarg\'o et al. 2003 for a more quantitative approach). 

Below we summarize the properties of these \ion{O}{6} wind absorption features:

(1) They are observed in 33\% of our sample (34\% in the MW, 32\% in the MC), in the spectra
of stars with spectral types from O7.5 to B0.5 and in every luminosity class. 
Since strong absorption components are preferentially found near
{\vinf} (see point 4 below) they are especially difficult to detect for
stars with $ 2400 \la v_\infty \la 3400$ \km\ owing to blending with the strongly
saturated interstellar \ion{H}{1} Ly$\beta$ line.
Stars with such terminal velocities range from O3 to about O7-O8 \citep{prinja90}.
Furthermore, these stars typically exhibit strongly saturated
P~Cygni profiles in the \ion{O}{6} resonance lines, which makes the
detection of localized absorption enhancements all the more difficult.
A similar problem arises when $v_\infty \la 600$ \km\ because of the contamination of the wing
of interstellar \ion{H}{1} Ly$\beta$ and \ion{C}{2} absorption lines.
Stars with such terminal velocities typically have spectral types later than B0.5 \citep{prinja90}.

(2) Their shapes are usually not Gaussian and can be asymmetric.

(3) They appear with different strengths, and their full-widths at half maximum vary approximately from less than 
100 \km\ up to 400--500 \km, the latter generally being observed near {\vinf} (see point 4).

(4) For the strong features, $v_{\rm abs}$ is similar to {\vinf} ($0.9 \,v_\infty \la v_{\rm abs} \la v_\infty$, 
except for Sk\,$-$67\degr05 where $v_{\rm abs} \approx 0.7 \, v_\infty$),
while for weak and medium features, $v_{\rm abs}$ is generally red-shifted by 
a few hundred \km\ with respect to {\vinf} ($0.7\, v_\infty \la v_{\rm abs} \la 0.9\, v_\infty$; except 
for Sk\,$-$67\degr101 where $v_{\rm abs} \approx 0.5 \, v_\infty$). 

(5) Some features appear to vary on time scales of a day, but other features are at the 
same velocity after several days or months have passed.

(6) The sparse time-series observations suggest that these features accelerate 
slowly with an acceleration of $\sim 10^{-3}$ to less than $\sim$$10^{-4}$ km\,s$^{-2}$ (see Figure~\ref{fig5} and
Figure~3 in Lehner et al. 2001). 

(7) There is no direct relation  between the widths of these features and 
the $v \sin i $ of star. Neither their width nor strength correlates with 
luminosity class or spectral type.

(8) There is no systematic difference in the frequency of occurence between the 
MW and MC sample, though
the size of the sample is certainly not large enough to draw any definitive
conclusions.

\section{Discussion and Summary}

We have described a survey for stellar wind variability 
based on sparse time-series observations of
a large sample of stars observed with {\fuse}.
This survey demonstrates that the winds in OB-type stars 
are variable in the
\ion{O}{6} doublet on various scales both in time 
and velocity. The typical signatures of wind variability 
are observed via (i) variation in the profiles for 56\% of the 
spectral types from O3 to B3 and 64\% of the spectral types from O3 to B1, 
and (ii) wind absorption features in 33\% of the cases for the whole sample, 
or 46\% of the cases for spectral types from O3 to B0.5. 

These percentages have to be considered lower limits. 
Owing to the nature of this survey, only a few exposures
separated  by a few days or a few months were obtained  
with no regard to the time scales believed to be relevant to
the variability of hot-star winds. 
Furthermore, the quality of {\fuse}\ data precludes the
detection of small-amplitude variations.
In view of the high incidence of variability detected
under these circumstances, variability in \ion{O}{6} wind
profiles is likely ubiquitous.

At spectral types later than B1, no \ion{O}{6} wind variability is observed. This
could be due to limitations in the detectability of wind activity 
caused by the weakness or absence of this feature in these stars,
as well as the smaller number of stars observed in those spectral types. 
In the former case, this might be explained by the fact that
for later spectral types, \ion{O}{3} becomes
the dominant ion, so that \ion{O}{6} cannot be produced 
via Auger ionization \citep{cassinelli79}.

The sample is not large enough to relate the frequency of the temporal 
variations or the absorption features to specific spectral types
or luminosity classes. But it {\em suggests} that stars 
with spectral types earlier than O8 may show variations
more often.

The observed variations extend over several hundreds of 
\km\ and can be strong. This favors the idea that the bulk of 
the X-rays producing \ion{O}{6} via Auger ionization 
originates from a few strong shocks. 

The width over which the wind profile of the \ion{O}{6} varies is poorly
correlated with the terminal velocity for the MW sample. 
There is, however,  a significant correlation 
(close to a 1:1 relationship) between the blue-edge 
velocity of the variation (i.e., at the maximum (negative)
velocity at which no variation 
is observed) and {\vinf} (see Figure~\ref{fig4}). 
This might imply that systematically  stronger shocks occur at larger
velocities (i.e., near {\vinf}).

Although the high velocity absorption features
seen in \ion{O}{6}
directly indicate the outflow of hot gas, it is not
clear from this survey exactly how these absorption 
features evolve with time. Figure~3 in \citet{lehner01}  
and Figure~\ref{fig5} give nice examples of velocity-varying 
\ion{O}{6} absorption wind features, but 
more intensive time-series observations are required to characterize
this evolution \citep[see][]{fullerton03}. 
We note, however, that the stronger (both in optical depth 
and width) components are near the terminal velocity, while 
weaker components appear and evolve at lower velocities.

By comparing the frequency of variability exhibited by
\ion{O}{6} to that seen in \ion{S}{4} and \ion{P}{5}, we found that only
a small percentage of the whole sample  shows
wind variations for the lower ions, 
and those variations are  mainly observed in supergiant stars. 
This may be largely due to two selection effects. One is that 
S and P are much less abundant than O, thereby making it more difficult to detect
variations in these ions. The other is that our sample is biased toward
late O-type and early B-type stars, where the wind becomes weaker for 
\ion{P}{5}; i.e., the \ion{P}{5} line becomes principally a photospheric line.

The low frequency of observed variability in \ion{S}{4} is harder  to 
understand, because \ion{S}{4} is expected to be the dominant ion in the
winds of late O- and early B-type stars. If it is not merely an abundance 
selection effect, then this result may imply that the highly ionized 
component of these cooler winds is enhanced compared with hotter winds. 
Suppose, e.g., that the variability exhibited by a wind line is 
proportional to its optical depth; i.e., $\Delta \tau \propto \tau$. 
The ionization fraction is $q \propto \tau/(f\lambda A_E)$ (see, e.g., Massa et al. 2002),
where $A_E$ is the abundance of the element. Since \ion{O}{6}
variability is detected more frequently than \ion{S}{4} variability, 
$\Delta \tau($\ion{O}{6}$) > \Delta \tau($\ion{S}{4}$)$, i.e. 
$ \tau($\ion{O}{6}$) >  \tau($\ion{S}{4}$)$, and 
$q($\ion{O}{6}$) > 0.01\, q($\ion{S}{4}$)$. If \ion{S}{4} is dominant, 
$q($\ion{S}{4}$) \sim 1$, and hence $q($\ion{O}{6}$) \ga 0.01$, which is at least
an order of magnitude larger than typically observed for early- 
and mid-O type stars \citep{massa02}. This evidence for an excessive
abundance of \ion{O}{6} in the winds of comparatively cool stars 
-- where it is least expected -- is consistent with the predictions of 
\citet{macfarlane94}, who concluded that the presence of X-rays
can substantially alter the ionization balance of lower-density winds. 

\acknowledgements
This work is based on data obtained
for the Guaranteed Time Team by the NASA-CNES-CSA FUSE mission operated 
by the Johns Hopkins University. Financial support to U. S.
participants has been provided by NASA contract NAS5-32985.
This research has made use of the NASA
Astrophysics Data System Abstract Service and the SIMBAD database,
operated at CDS, Strasbourg, France.
We thank the referee, Doug Gies, for
helpful comments that improved the presentation of these results.

\newpage

\begin{figure*}[!th]
\begin{center}
\includegraphics[height=21. truecm]{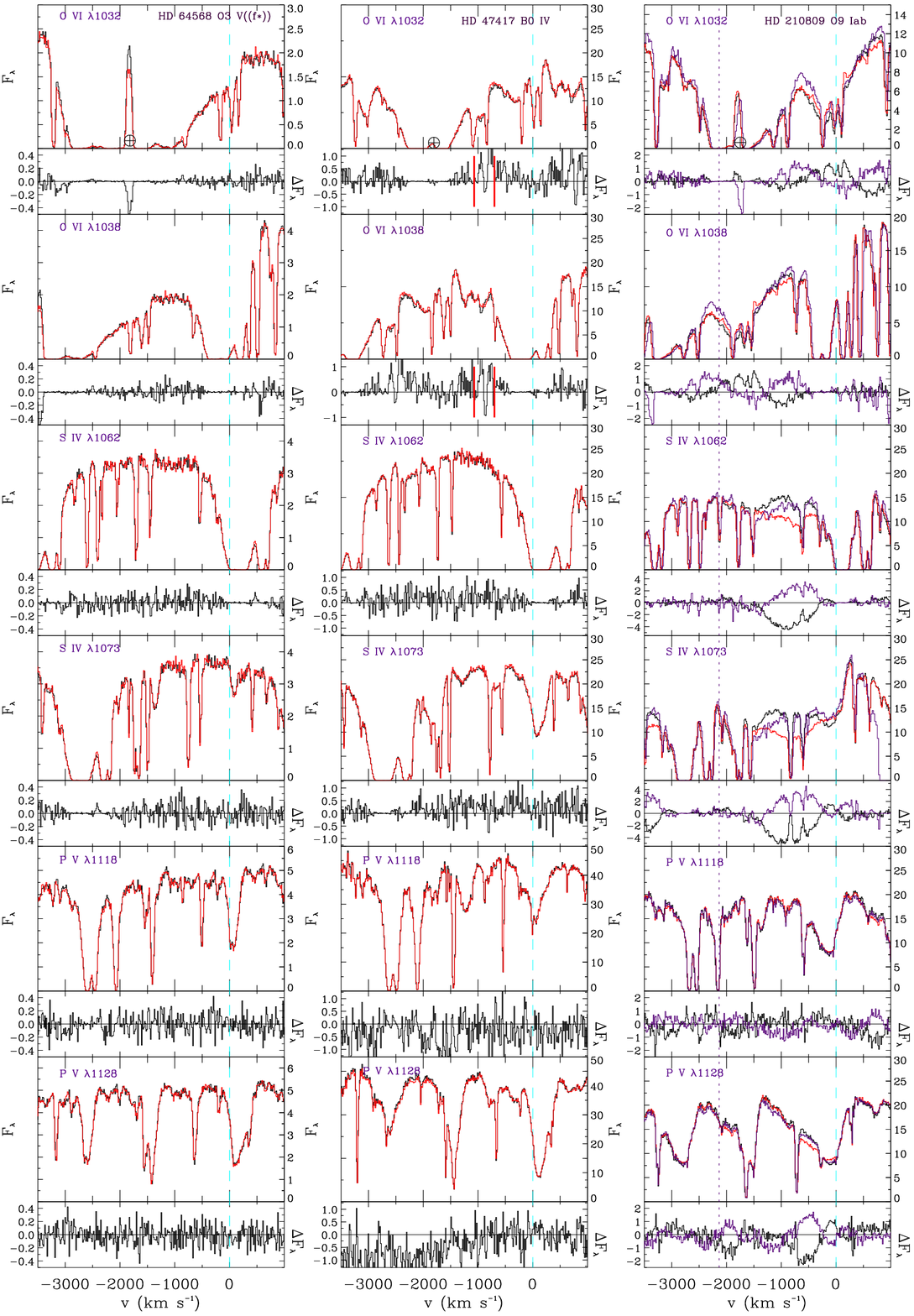}
\caption{}
\label{fig1}
\end{center}
\end{figure*}

FIGURE CAPTIONS:
\smallskip

Figure~\ref{fig1}: 
  Examples of {\fuse} spectra for three Galactic stars
  for the \ion{O}{6} $\lambda\lambda$1032, 1038,
  \ion{S}{4} $\lambda\lambda$1062, 1073, and 
  \ion{P}{5} $\lambda\lambda$1118, 1128 doublets. 
  The spectra and difference spectra are shown in units of
  $10^{-12}$ erg\,cm$^{-2}$\,s$^{-1}$\,\AA$^{-1}$. 
  For HD\,64568 (left panel), the observations were separated by 2.5 days; 
  for HD\,47417 (middle panel), the observations were separated by 0.8 day,
  and for HD\,210509 (right panel), the observations were separated by 1.4 and 1.3 days.
  The black, red, and blue profiles denote the first, second, and third observations, respectively. 
  The  dotted line in the right panel indicates the terminal velocity of 
  HD\,210509. 
   The position of the Lyman $\beta $ airglow line is indicated in the
  uppermost panels. The range of variability in the \ion{O}{6} line of
  HD\,47417 is emphasized
  by vertical red lines in difference flux panel.

Figure~\ref{fig2}: 
  The logarithm of the ratio of the fluxes 
  (taken at different times indicated by the $\Delta t$ in days) 
   vs.  velocity for the {\ion{O}{6}} $\lambda\lambda$1032, 1038,
  \ion{S}{4} $\lambda\lambda$1062, 1073, and 
  \ion{P}{5} $\lambda\lambda$1118, 1128 for 4 stars in the Galaxy and the LMC, showing the variety of 
  variations in the profiles. Note that the $y$-scale of a panel is the same for a given
  species but can change from species to species.
  Vertical dotted lines indicate {\vinf}, while the extent of the variation is 
  shown by the solid thick vertical lines. The variation in the profile ratio of {\ion{S}{4}}
  and \ion{P}{5} in the spectra of HD\,187459 at $\sim 0$ \km\ 
  indicates radial velocity variations, perhaps due to motion in
  a binary system.

Figure~\ref{fig3}: 
Fraction of stars in the combined Galactic and Magellanic Clouds sample
exhibiting \ion{O}{6} wind variability for different temperature (left panel)
or luminosity classes (right panel).  The dotted line
represents the average  fraction of observed \ion{O}{6} wind variability.

Figure~\ref{fig4}: 
The full-width ($\Delta v$, top panel) and blue edge ($v_{\rm max}$, bottom panel) of the variability
observed in the \ion{O}{6} wind with respect to the terminal velocity ($v_\infty$).
Triangles correspond to the MW stars, and squares denote stars in the LMC.
Typical measurement errors are about 10\% for {\vinf} and 100--200 \km\ for $\Delta v$
and $v_{\rm max}$. The dotted lines indicate 
a 1:1 relationship. The dashed lines are a least-squares fit  weighted by the inverse 
of the quadratic sum of the errors  to the data 
(only MW sample in top panel, both samples in bottom panel).

Figure~\ref{fig5}: 
\ion{O}{6}  absorption components in the wind profiles
of early-type stars in the Galaxy. 
The profiles are plotted in flux units versus velocity with
respect to \ion{O}{6} $\lambda$1032 (panel a) or $\lambda$1038 (panel b).
$\Delta t$ indicates the time in days between successive
observations. Black, red, and blue profiles indicate the first, second, and
third observations, respectively.
The vertical dotted lines indicate the ``centroid'' of the \ion{O}{6} DAC.
See \citet{lehner01} for other examples in the MW and the LMC.

\newpage

\begin{figure*}[!th]
\begin{center}
\includegraphics[height=21. truecm]{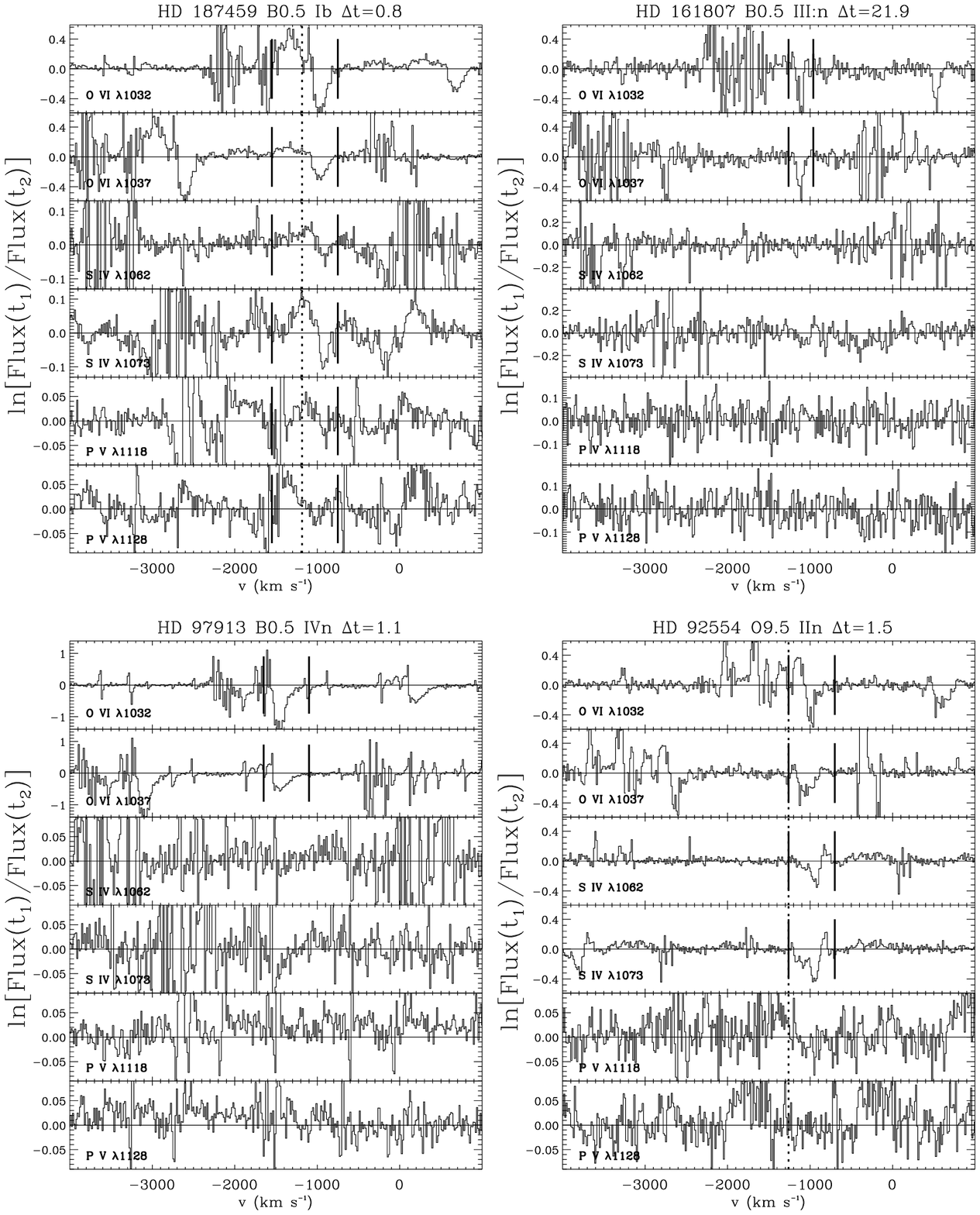}
\caption{} 
\label{fig2}
\end{center}
\end{figure*}

\begin{figure*}[!th]
\begin{center}
\includegraphics[height=21. truecm]{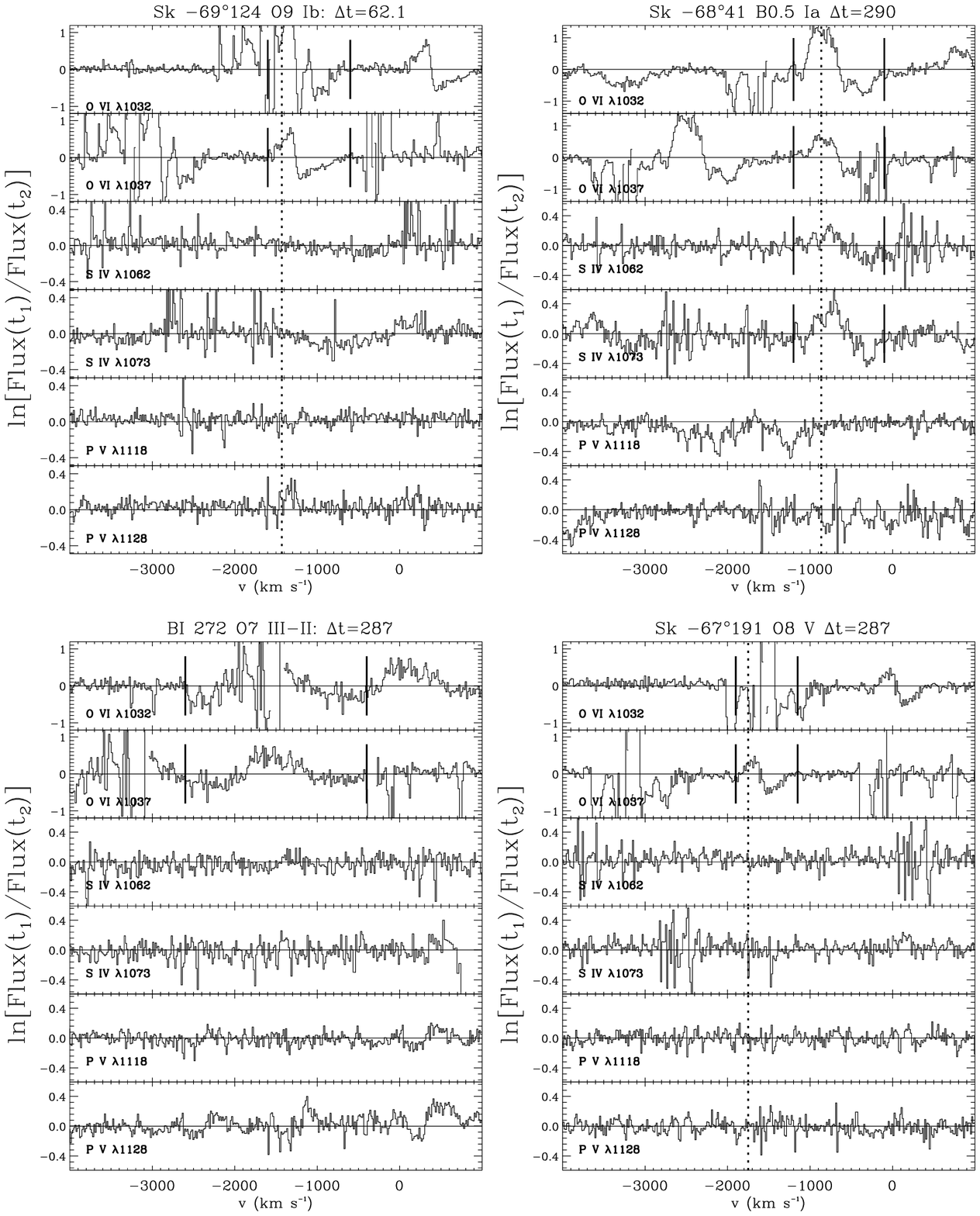}
\figurenum{\ref{fig2}}
\caption{Continued. LMC stars.} 
\end{center}
\end{figure*}

\begin{figure*}[!th]
\begin{center}
\includegraphics[width=18.truecm]{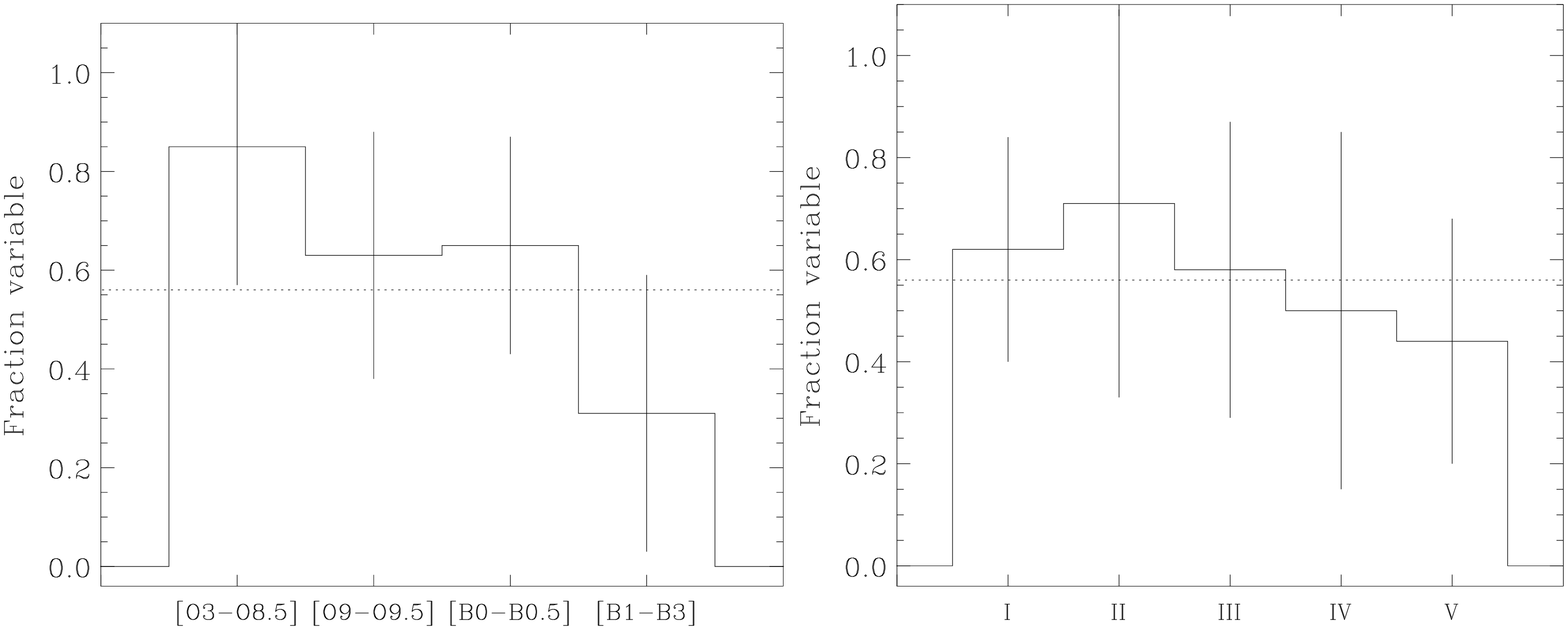}
\caption{} 
\label{fig3}
\end{center}
\end{figure*}

\begin{figure*}[!th]
\begin{center}
\includegraphics[width=12.5 truecm]{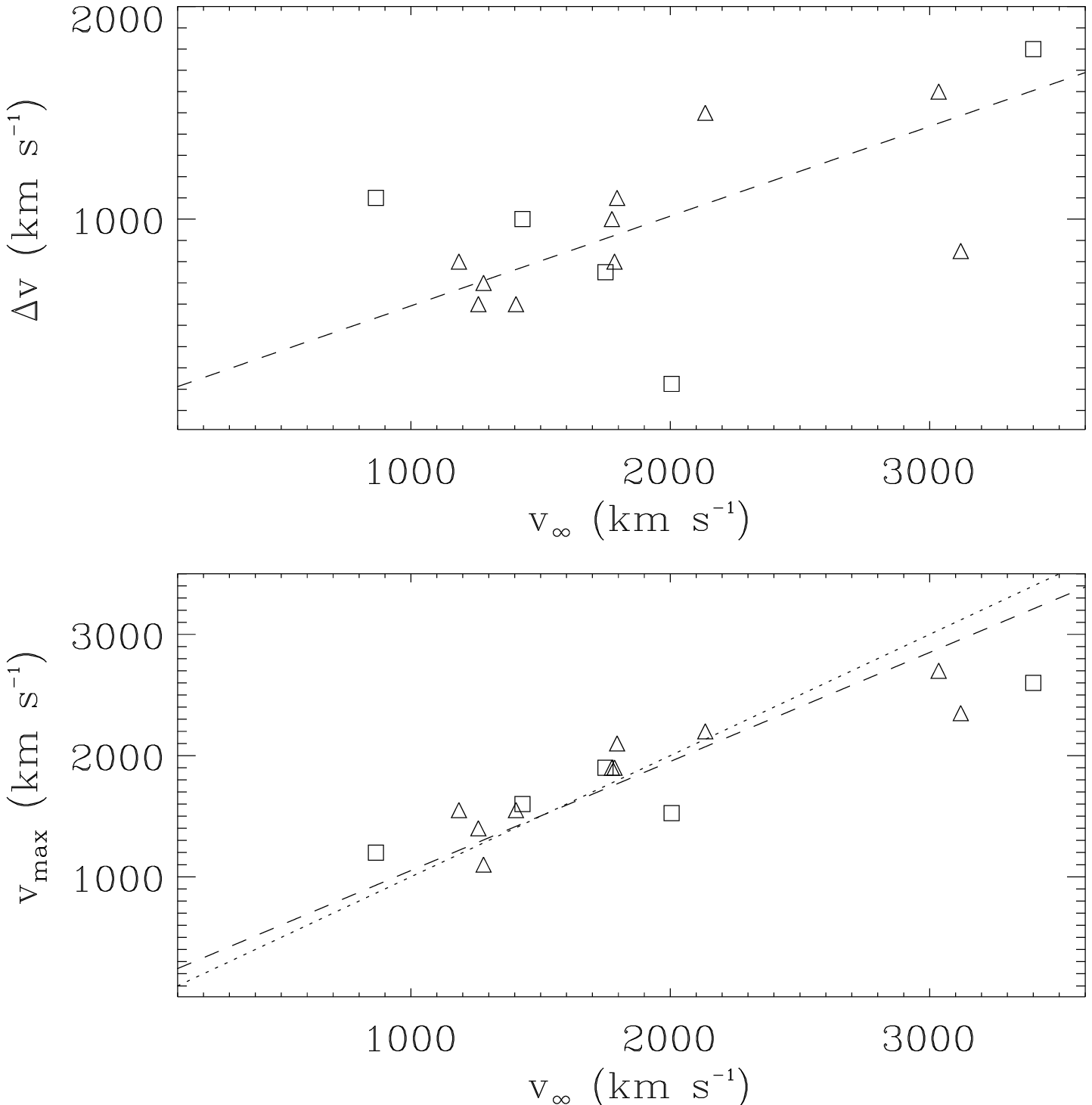}
\caption{}
\label{fig4}
\end{center}
\end{figure*}

\begin{figure*}[!th]
\begin{center}
\includegraphics[height=21. truecm]{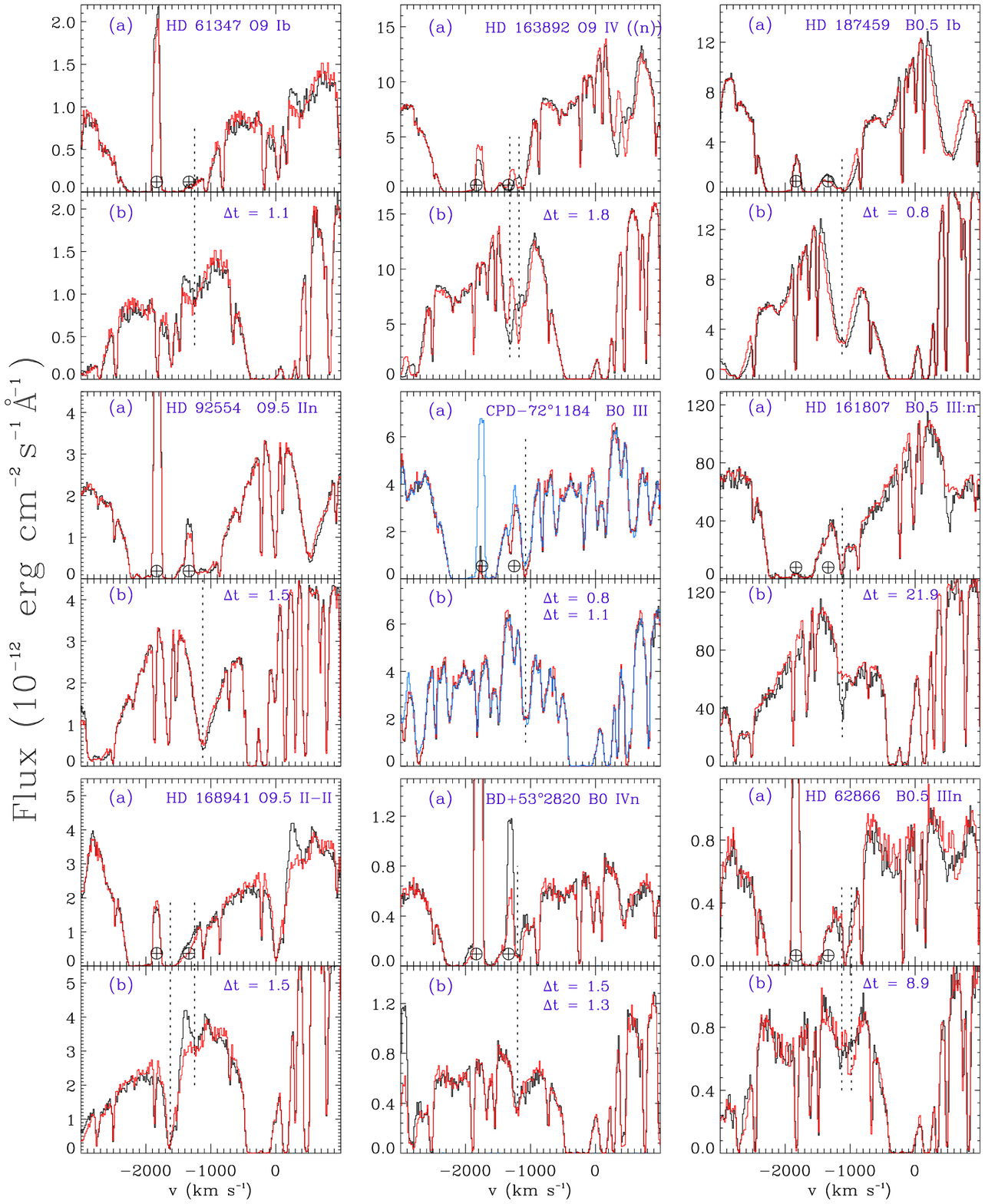}
\caption{}
\label{fig5}
\end{center}
\end{figure*}

\newpage
\begin{deluxetable}{llrrccccccc}
\tablecolumns{8}
\tablewidth{0pc} 
\tabletypesize{\scriptsize}
\tablecaption{Observations of Galactic Stars \label{t1}} 
\tablehead{\colhead{Star}    &   \colhead{Sp. Type}&   \colhead{$l$}&   \colhead{$b$}  &  \colhead{$V$}&  \colhead{$v\sin i$} &\colhead{$\Delta t$\tablenotemark{a}}&\colhead{UT-Date} &\colhead{Rootname}&\colhead{Remarks\tablenotemark{b}} \\
\colhead{}    &   \colhead{Reference}&   \colhead{(\degr)}&   \colhead{(\degr)}  &  \colhead{$E(B-V)$}&  \colhead{$v_\infty$} &\colhead{(day)} &\colhead{}&\colhead{}\\
\colhead{}    &   \colhead{}&   \colhead{}&   \colhead{}  &  \colhead{(mag.)}&  \colhead{(\km)} &\colhead{} &\colhead{} &\colhead{}&\colhead{}}
\startdata
HDE\,303308     &  O3\,V((f))   & 287.60   & 	$-0.61$	    &	  8.17	&   111$^b$	  & 1.7	&     2000-05-25 &    P1221601      &	      \\
	        &  W73	        &   	   &		    &	0.45	&  3035$^b$	  &  	&     2000-05-27 &    P1221602      &	      \\
HD\,64568       & O3\,V((f$^*$))& 243.14   &	$+0.71$     &	  9.38  &  \nodata 	  & 2.5	&     2000-03-25 &    P1221104      &	    \\
	        &  W82	        &   	   &		    &	 0.37 	&  \nodata	  & 5.7 &     2000-04-03 &    P1221102      &	    \\
	        &  	        &   	   &		    &		&	  	  &  	&     2000-04-05 &    P1221103      &	    \\
CPD-59\degr2600	&  O6\,V((f))   & 287.60   &	$-0.74$	    &	  8.61  &  142$^b$  	  & 1.9	&     2000-03-21 &    P1221401      &	      \\
	        &  W73	        &   	   &		    &	0.53 	&  3120$^b$   	  & 1.9	&     2000-03-23 &    P1221402      &	      \\
	        &  	        &   	   &		    &		&	  	  &  	&     2000-03-25 &    P1221403      &	      \\
CPD-59\degr2603	&  O7\,V        & 202.18   &  $-58.98$      &	  8.77  & 164$^{b,\alpha}$& 67.3&     2000-01-30 &    S3040501      &	      \\
	        &  W73	        &   	   &		    &	0.46	&1840$^{b,\alpha}$& 17.0&     2000-03-21 &    P1221501      &	      \\
	        &  	        &   	   &		    &		&	  	  &  	&     2000-04-06 &    S3040502      &	      \\
HD\,152590      &  O7.5\,V      &  344.84  &	$+1.83$     &	8.48    &    60$^b$    	  & 35.1&     2001-07-08 &    B0710601      &RV var   \\
	        &   W72	        &   	   &		    &	0.38    & 1785$^b$        &  	&     2001-07-12 &    B0710602      &	      \\
HD\,61347       &  O9\,Ib       & 230.60   &	$+3.79$     &	  8.43  & 116$^b$	  & 1.1	&     2000-04-15 &    P1022001      &	      \\
	        &  M55	        &   	   &		    &	0.45	&  1775$^b$       &  	&     2000-04-16 &    P1022002      &	      \\
HD\,210809      &  O9\,Iab      & 99.85    &	$-3.13$	    &	  7.54  &   117$^b$   	  & 1.4	&     2000-08-05 &    P1223101      &RV var   \\
	        &  W76	        &   	   &		    &	0.33	&  2135$^b$  	  & 1.3	&     2000-08-07 &    P1223102      &	      \\
	        &  	        &   	   &		    &		&	  	  &  	&     2000-08-08 &    P1223103      &	      \\
HD\,153426      &  O9\,II       & 258.07   &  $-15.37$      &     7.47  &  108$^b$        &105  &     2000-03-31 &    P1027201      &RV var   \\
	        &  W73	        &   	   &		    &	 0.45	&  2200$^b$  	  &  	&     2000-07-14 &    P1027202      &	      \\
HD\,91651       &  O9\,V:n      & 286.55   &	$-1.72$	    &	  8.84  &  292$^b$	  & 1.3	&     2000-05-25 &    P1023101      &	      \\
	        &  W73	        &   	   &		    &	0.29	&  1705$^b$   	  &  	&     2000-05-27 &    P1023102      &	      \\
HD\,92554       &  O9.5\,IIn    & 287.60   &	$-2.02$	    &	  9.47  &   298$^b$	  & 1.5	&     2000-05-26 &    P1023201      &RV var?  \\
	        &  G77	        &   	   &		    &	0.39	& 1260$^b$  	  &  	&     2000-05-27 &    P1023202      &	      \\
HD\,168941      &  O9.5\,II-III & 275.23   &    $-3.62$     &     9.34  &   116$^b$	  & 2.2	&     2000-08-30 &    P1016501      &RV var   \\
	        &  W82	        &   	   &		    &	 0.37 	&  1795$^b$  	  &  	&     2000-09-01 &    P1016502      &	      \\
HD\,156292      &  O9.5\,III    & 345.35   &    $-3.08$     &     7.49  &   102$^b$	  &1.9  &     2000-04-03 &    P1027402      &RV var   \\
	        &  W73	        &   	   &		    &	 0.56 	&   1355$^b$	  &  	&     2000-04-05 &    P1027403      &	      \\
HD\,163892      &  O9\,IV ((n)) & 7.15     &	$+0.62$     &	7.44    &   201$^b$       & 1.8	&     2001-04-27 &    P1027601	    &RV var   \\
	        &  W73 	        &   	   &		    &	0.47    & 1405$^b$        &  	&     2001-04-29 &    P1027602	    &	      \\
HDE\,308813     &  O9.5\,V      & 294.80   &	$-1.61$	    &	  9.38  &  \nodata    	  & 2.0	&     2000-03-23 &    P1221901      &RV var   \\
		&  S70	        &   	   &		    &	0.37	&\nodata	  & 1.3	&     2000-03-25 &    P1221902      &	      \\
		&	        &   	   &		    &		&	  	  &  	&     2000-03-27 &    P1221903      &	      \\
CPD$-$72\degr1184& B0\,III      & 233.12   &   $-61.63$     &    10.68  &  \nodata 	  & 0.8	&     2000-03-27 &    S5140101      &	      \\
		&  H74	        &   	   &		    &	0.23 	&\nodata	  & 1.1	&     2000-03-28 &    S5140102      &	      \\
		&	        &   	   &		    &		&	  	  &  	&     2000-03-29 &    S5140103      &	      \\
HD\,192035	&  B0\,III-IV(n)& 68.81    &	$+3.85$	    &	  6.44  &  155$^c$  	  & 2.5	&     2000-06-17 &    P1028601      &	      \\
		&  W71	        &   	   &		    &	0.42	&\nodata	  & 2.0	&     2000-06-19 &    P1028602      &	      \\
		&   	        &   	   &		    &	 	&       	  &  	&     2000-06-21 &    P1028603      &	      \\
HD\,47417	&  B0\,IV       & 205.35   &	$+0.35$     &	  6.97  &  \nodata	  & 0.8	&     2000-03-15 &    P1021602      &	      \\
		&  M55	        &   	   &		    &	0.31	&\nodata	  &  	&     2000-03-16 &    P1021601      &	      \\
HD\,203374A     &  B0\,IVpe     & 100.26   &	$+8.58$     &	6.69    &  350$^c$        & 0.4 &     2001-08-02 &    B0300101      &	      \\
	        &  M55 	        &   	   &		    &	0.60    & \nodata         &  	&     2001-08-02 &    B0300102      &	      \\
BD\,+53\degr2820&  B0\,IVn      & 9.26     &   $+58.16$     &     9.95  &  \nodata	  & 1.5	&     2000-08-06 &    P1223201      &	      \\
		&  H56	        &   	   &		    &	0.40 	&\nodata 	  & 1.3	&     2000-08-07 &    P1223202      &	      \\
		&   	        &   	   &		    &	 	&       	  &  	&     2000-08-08 &    P1223203      &	      \\
HD\,191495	&  B0\,IV-V     & 72.74    &	$+1.41$     &	  8.26  &  235$^c$ 	  & 1.0	&     2000-08-10 &    P1222901      &RV var   \\
		&  W71	        &   	   &		    &	0.40 	&\nodata  	  &  	&     2000-08-11 &    P1222902      &	      \\
HD\,195965	&  B0\,V        & 83.33    &	$+7.76$     &	  8.20  &  \nodata	  &141.2&     2000-06-17 &    P1028801      &	      \\
		&  M55	        &   	   &		    &	 0.36	&\nodata  	  & 0.9 &     2000-11-08 &    P1028802      &	      \\
		&	        &   	   &		    &		&	  	  &  	&     2000-11-09 &    P1028803      &	      \\
HD\,186994      &   B0.2IV      & 78.62    &	$+10.06$    &	7.51    &  125$^c$     	  &63.4 &     2001-07-02 &    P2160801      &RV var   \\
	        &   W71         &   	   &		    &	0.20    &  \nodata        &  	&     2001-09-07 &    P2160802      &	      \\
HD\,207538      &  B0.2\,V      & 12.12    &  $+64.38$      &     7.31  &   51$^b$ 	  &226  &     1999-12-08 &    P1162902      & 	      \\
	        &  MA02	        &   	   &		    &	0.63 	&\nodata	  &  	&     2000-07-21 &    P1162903      &	      \\
HD\,187459	&  B0.5\,Ib     & 68.81    &	$+3.85$	    &	  6.44  & 139$^b$	  & 0.8	&     2000-08-10 &    P1028201      &RV var?  \\
		&  M55	        &   	   &		    &	 0.42	&1185$^b$	  &  	&     2000-08-11 &    P1028202      &	      \\
HD\,224151	&  B0.5\,II     & 115.44   &	$-4.64$	    &	  6.00  &  115$^b$     	  &2.1  &     2000-08-11 &    P1224101      &RV var   \\
		&  M55	        &   	   &		    &	0.48	&1280$^b$   	  &0.6  &     2000-08-13 &    P1224102      &	      \\
		&	        &   	   &		    &		&	  	  &  	&     2000-08-14 &    P1224103      &	      \\
HDE\,332407	&  B0.5\,III:   & 64.28    &	$+3.11$	    &	  8.50  &  140$^c$ 	  & 1.8	&     2000-06-10 &    P1222801      &	      \\
		&  W71	        &   	   &		    &	 0.48	& \nodata	  &  	&     2000-06-12 &    P1222802      &	      \\
HD\,161807	&  B0.5\,III:n  & 351.78   &   $-5.85 $     &	  6.99  &  350$^c$   	  & 21.9&     2000-08-17 &    P1222301      & 	      \\
		&  G77	        &   	   &		    &	0.23 	&\nodata  	  &  	&     2000-09-08 &    P1222302      &	      \\
HD\,172140	&  B0.5\,III    &  5.28    &   $-10.61$     &	  9.93  & \nodata   	  & 1.5	&     2000-05-18 &    P1016601      &	      \\
		&  H70	        &   	   &		    &	0.25 	&\nodata  	  & 2.1	&     2000-05-20 &    P1016602      &	      \\
		&	        &   	   &		    &		&	  	  &  	&     2000-05-22 &    P1016603      &	      \\
HD\,62866	&  B0.5\,IIIn   & 237.48   &	$+1.80$     &	  9.01  &  \nodata	  & 8.9	&     2000-04-02 &    P1221002      &	      \\
		&  G77	        &   	   &		    &	0.35	&\nodata  	  &  	&     2000-04-11 &    P1221004      &	      \\
HD\,97913	&  B0.5\,IVn    & 290.84   &	$+1.41$     &	  8.80  &   310$^c$	  & 1.1	&     2000-05-26 &    P1221701      &	      \\
		&  G77	        &   	   &		    &	0.32	&\nodata   	  &  	&     2000-05-27 &    P1221702      &	      \\
BD\,+35\degr4258&  B0.5\,Vn     & 77.19    &	$-4.74$	    &	 9.41   &  \nodata 	  & 3.0	&     2000-06-14 &    P1017901      &RV var   \\
		&  M55	        &   	   &		    &	0.29	&\nodata 	  &  	&     2000-06-17 &    P1017902      &	      \\
HD\,148422	&  B1\,Ia       & 329.92   &	$-5.60$	    &	  8.60  &   81$^b$    	  & 2.1	&     2000-04-08 &    P1015001      &	      \\
		&  HC75	        &   	   &		    &	0.28	&1335$^b$	  & 1.9	&     2000-04-10 &    P1015002      &	      \\
		&	        &   	   &		    &		&	  	  &  	&     2000-04-12 &    P1015003      &	      \\
HD\,191877	&  B1\,Ib       & 61.57    &	$-6.48$	    &	  6.28  &  152$^b$  	  & 420 &     2000-06-05 &    P1028701      &RV var?  \\
		&  M55	        &   	   &		    &	0.18	&1160$^b$  	  & 	&     2001-07-30 &    P2051101      &	      \\
HDE\,235783	&  B1\,Ib       & 101.69   &	$-1.87$	    &	  8.68  &  93$^b$  	  & 1.5	&     2000-08-05 &    P1223301      &	      \\
		&  M55	        &   	   &		    &	0.36	&1070$^b$  	  & 1.3	&     2000-08-07 &    P1223302      &	      \\
		&	        &   	   &		    &		&	  	  &  	&     2000-08-08 &    P1223303      &	      \\
BD\,+48\degr3437&  B1\,Iab      & 354.70   &   $+58.01$     &     8.69  &  54$^b$	  & 1.3	&     2000-08-06 &    P1018401      &	      \\
		&  M55	        &   	   &		    &	0.35	&\nodata   	  & 1.4	&     2000-08-07 &    P1018402      &	      \\
		&	        &   	   &		    &		&	  	  &  	&     2000-08-09 &    P1018403      &	      \\
HD\,225757	&  B1\,IIIn     & 309.48   &   $+54.50$     &    10.59  &  \nodata 	  & 2.0	&     2000-08-06 &    P1017701      &	      \\
		&  C76	        &   	   &		    &	0.22	&\nodata   	  & 0.6	&     2000-08-08 &    P1017702      &	      \\
		&	        &   	   &		    &		&	  	  &  	&     2000-08-09 &    P1017703      &	      \\
HD\,91597	&  B1\,IIIne    & 286.86   &	$-2.37$	    &	 9.84   &  \nodata   	  &  1.4&     2000-02-04 &    P1023002      &	      \\
		&  G77	        &   	   &		    &	0.30	&\nodata   	  & 2.3	&     2000-02-06 &    P1023003      &	      \\
		&   	        &   	   &		    &	  	&        	  &  	&     2000-02-08 &    P1023004      &	      \\
BD\,+52\degr3210&  B1\,V    	&10.9608   &   $+56.31$     &   10.69   & \nodata 	  & 1.3	&     2000-08-06 &    P1223501      &	      \\
		&  M55	        &   	   &		    &	 0.24	&\nodata   	  & 1.3	&     2000-08-07 &    P1223502      &	      \\
		&   	        &   	   &		    &	  	&       	  &  	&     2000-08-08 &    P1223503      &	      \\
HD\,73   	&  B1.5\,IV     & 114.17   &    $-18.69$    &    8.48   &  105$^c$ 	  &49.9	&     2000-08-01 &    P1010101      & 	      \\
		&  W71	        &   	   &		    &	0.07 	&\nodata   	  &  	&     2000-09-20 &    P1010102      &	      \\
HD\,202347	&  B1.5\,V      & 44.85    &   $+57.17$     &    7.50   &  125$^c$	  &84.7	&     2000-06-20 &    P1028901      &	      \\
		&  W71	        &   	   &		    &	0.17 	&\nodata  	  &  	&     2000-09-13 &    P1028902      &	      \\
BD\,+53\degr2885&  B2\,III      & 102.75   &	$-2.93$	    &	 10.46  &  \nodata   	  & 2.7	&     2000-08-02 &    P1223601      &	      \\
		&  M55	        &   	   &		    &	 0.28	&\nodata   	  &  	&     2000-08-05 &    P1223602      &	      \\
HD\,47961       &  B2\,V        & 203.02   &	$+2.28$     &	7.50    &    \nodata      & 11.7&     2001-02-20 &    P1310401      &	      \\
	        &  K69  	&   	   &		    &	0.08    & \nodata	  &  	&     2001-03-04 &    P1310402      &	      \\
HD\,51013       &  B3\,V        & 235.11   &	$-10.13$    &	8.81    &   \nodata       & 438 &     2000-01-24 &    A0630901	    &	      \\
	        &  HS88         &   	   &		    &	0.05    & \nodata	  &  	&     2001-04-05 &    A0630902	    &	      \\
HD\,47777       &  B3\,V        & 203.12   &	$+2.03$     &	7.90    &    135$^c$   	  & 11.7&     2001-02-20 &    P1310201      &	      \\
	        &  G68 	        &   	   &		    &	0.07    & \nodata	  &  	&     2001-03-04 &    P1310202      &	      \\
\enddata
\tablenotetext{a}{$\Delta t$, time between successive observations.}
\tablenotetext{b}{Radial velocity variation detected or suspected.}
\tablerefs{($a$) Sources of spectral type: 
C76$=$\citet{crampton76};
G77$=$\citet{garrison77};
G68$=$\citet{guetter68};
H56$=$\citet{hiltner56};
H70$=$\citet{hill70};
H74$=$\citet{hill74};
HC75$=$\citet{houck75};
HS88$=$\citet{houck88};
K69$=$\citet{karlsson69};
M55$=$\citet{morgan55};
MA02$=$\citet{maiz02};
S70$=$\citet{schild70};
W71, W72, W73, W76, W82$=$\citet{walborn71,walborn72,walborn73,walborn76,walborn82}.
($b$) $v_\infty$ from \citet{prinja90,howarth97}; $v\sin i $ from \citet{howarth97} but see also 
\citet{penny}.
($c$) $v\sin i $ from \citet{uesugi}. ($\alpha$) CPD$-$59\degr2603 is probably a multicomponent and only parameters for component A
\citep{howarth97} are indicated here.}
\end{deluxetable}

\newpage
\begin{deluxetable}{llrrccccccc}
\tablecolumns{8}
\tablewidth{0pc} 
\tabletypesize{\scriptsize}
\tablecaption{Observations of Stars in the Magellanic clouds\tablenotemark{a} \label{t2}} 
\tablehead{\colhead{Star}    &   \colhead{Sp. Type}&   \colhead{$l$}&   \colhead{$b$}  &  \colhead{$V$}&  \colhead{$v_\infty$} &\colhead{$\Delta t$\tablenotemark{b}}&\colhead{UT-Date} &\colhead{Rootname}&\colhead{Remarks\tablenotemark{c}} \\
\colhead{}    &   \colhead{Reference}&   \colhead{(\degr)}&   \colhead{(\degr)}  &  \colhead{$E(B-V)$}&  \colhead{(\km)} &\colhead{(day)} &\colhead{}&\colhead{}\\
\colhead{}    &   \colhead{}&   \colhead{}&   \colhead{}  &  \colhead{(mag.)}&  \colhead{} &\colhead{} &\colhead{} &\colhead{}&\colhead{}}
\startdata
Sk\,$-$67\degr167	&  O4\,Inf$^+$   & 277.87   &    $-32.47$    & 12.54	&  2005$^b$	&285	&  1999-12-17 &    P1171902  &	      \\
	       		&W95	         &	    &		     &	0.14	&		&	&  2000-09-27 &    P1171901  &	      \\
Sk\,$-$71\degr45	&  O4-5\,IIIf    & 281.86   &  $-32.03$      & 11.47	&  2500$^c$	&  2.7	&  2000-10-02 &    P1031501  &	      \\
	       		&W77	         &	    &		     &	0.20	&		&  1.0	&  2000-10-05 &    P1031503  &	      \\
	       		&	         &	    &		     &	        &		&  0.5	&  2000-10-06 &    P1031502  &	      \\
	       		&	         &	    &		     &	        &		&	&  2000-10-06 &    P1031504  &	      \\
Sk\,$-$67\degr111	& O6\,Ia(n)fp var& 277.75   &  $-32.97$      & 12.57	&  2090$^d$	& 138	&  1999-09-26 &    X0200101  &RV var? \\
	       		& W02	         &	    &		     &	0.11	&		&	&  2000-10-20 &    P1173001  &	      \\
BI\,208			&  O7\,V$z$      & 277.54   &  $-32.30$      & 14.02    &  \nodata	&  0.5	&  2000-09-30 &    P1172704  &	      \\
	       		& Wpc	         &	    &		     &	0.03    &		&  0.1	&  2000-10-01 &    P1172703  &	      \\
	       		&	         &	    &		     &	       	&		&  1.2	&  2000-10-01 &    P1172702  &	      \\
	       		&	         &	    &		     &	       	&		&	&  2000-10-01 &    P1172705  &	      \\
BI\,272		        &  O7\,III-II:   & 277.24   &  $-31.32$      & 13.20	&  3400$^c$	& 287	&  1999-12-18 &    P1172902  &	      \\
	       		&Wpc	         &	    &		     &	0.17	&		&	&  2000-09-29 &    P1172901  &	      \\
Sk\,$-$67\degr101	&  O8\,II((f))   & 277.78   &  $-33.05$      & 12.63   	&  2005$^b$	& 284	&  1999-12-20 &    P1173403  &	      \\
	       		& W02	         &	    &		     &		&		&	&  2000-09-29 &    P1173401  &	      \\
BI\,173		        &  O8\,II:       & 279.67   &  $-32.68$      & 13.00	&  2850$^c$	&  61.9	&  2000-10-03 &    P1173201  &RV var  \\
	       		& W02	         &	    &		     &	0.17	&		&	&  2000-12-04 &    P1173202  &	      \\
Sk\,$-$67\degr191	&  O8\,V         & 277.66   &  $-32.33$      & 13.46	&   1750$^b$	& 287	&  1999-12-17 &    P1173102  &	      \\
	       		&C86	         &	    &		     &	0.10    &		&	&  2000-09-29 &    P1173101  &	      \\
Sk\,$-$68\degr03	&  O9\,I         & 279.70   &  $-35.93$      & 13.13	& \nodata	&  1.3	&  2000-10-04 &    A0490402  &	      \\
	       		& C86	         &	    &		     &	0.48 	&		&	&  2000-10-05 &    A0490401  &	      \\
Sk\,$-$69\degr124	&  O9\,Ib:       & 279.61   &  $-32.86$      & 12.66	&  1430$^b$	&  62.1	&  2000-10-03 &    P1173601  &	      \\
	       		&Wpc	         &	    &		     &	 0.12	&		&	&  2000-12-04 &    P1173602  &	      \\
Sk\,$-$65\degr44	&  O9\,V         & 277.78   &  $-33.05$      & 12.63	& \nodata      	&  140	&  2001-10-27 &    P1173401  &	      \\
	       		& C86	         &	    &		     &	0.14  	&		&	&  2002-03-16 &    P1173401  &	      \\
Sk\,$-$67\degr05	&  O9.7\,Ib      & 278.89   &  $-36.32$      & 11.34	& 1665$^d$	&  2.8	&  2000-10-04 &    P1030704  &	      \\
	       		&F88	         &	    &		     &	0.15	&		&	&  2000-10-07 &    P1030703  &	      \\
Sk\,$-$65\degr21	&  O9.7\,Iab     & 276.19   &  $-35.79$      & 12.02	& 1330$^d$	& 0.6	&  2000-10-05 &    P1030901  &	      \\
	       		&W95	         &	    &		     &	 0.20	&		& 0.1	&  2000-10-05 &    P1030904  &	      \\
	       		&	         &	    &		     &		&		& 0.4	&  2000-10-05 &    P1030903  &	      \\
	       		&	         &	    &		     &		&		&	&  2000-10-05 &    P1030902  &	      \\
Sk\,$-$71\degr08	&  O9\,II        & 282.52   &  $-33.88$      & 13.25	& \nodata  	&  636	&  1999-12-21 &    A0491401  &	      \\
	       		& C86	         &	    &		     &	0.08 	&		&	&  2001-09-17 &    A0491402  &	      \\
Sk\,$-$70\degr85	&  B0\,I         & 281.26   &  $-33.31$      & 12.30 	& \nodata      	&  636	&  1999-12-21 &    A0491301  &	      \\
	       		& J01	         &	    &		     &	0.15  	&		&	&  2001-09-17 &    A0491302  &	      \\
Sk\,$-$68\degr41	&  B0.5\,Ia      & 279.02   &  $-34.82$      & 12.01	& 865$^b$	& 290	&  1999-12-18 &    P1174102  &	      \\
	       		&F91	         &	    &		     &	0.16	&		&	&  2000-10-03 &    P1174101  &	      \\
AV\,488			&  B0.5\,Iaw     & 300.51   &  $-43.66$      & 11.90	&1040$^b$	& 0.1	&  2000-10-08 &    P1176803  &	      \\
	       		&W95	         &	    &		     &	0.14	&		& 2.0	&  2000-10-08 &    P1176802  &	      \\
	       		&	         &	    &		     &		&		&	&  2000-10-10 &    P1176801  &	      \\
Sk\,$-$67\degr28	&  B0.7\,Ia      & 278.06   &  $-35.69$      & 12.28 	&  \nodata      &  4.1	&  1999-12-16 &    A0490201  &	      \\
	       		& F88	         &	    &		     &	0.10  	&		&	&  1999-12-20 &    A0490202  &	      \\
Sk\,$-$68\degr75	&  B1\,I         & 278.65   &  $-33.17$      & 12.03	&  \nodata     	&  6.5	&  2000-10-03 &    A0490501  &	      \\
	       		& J01	         &	    &		     &	0.19 	&		&	&  2000-10-10 &    A0490502  &	      \\
Sk\,$-$70\degr120	&  B1\,Ia        & 280.72   &  $-30.47$      & 11.59	& \nodata      	&  0.1	&  2000-09-28 &    A0491002  &	      \\
	       		& F88	         &	    &		     &	0.14  	&		&	&  2000-09-28 &    A0491001  &	      \\
Sk\,$-$67\degr14	&  B1.5\,Ia      & 278.27   &  $-36.05$      & 11.52	& 610$^b$	&  4.3	&  2000-09-27 &    P1174201  &	      \\
	       		&F91	         &	    &		     &	 0.10	&		&  0.6	&  2000-10-01 &    P1174203  &	      \\
	       		&	         &	    &		     &		&		&	&  2000-10-02 &    P1174202  &	      \\
AV\,18			&  B2\,Ia        & 303.36   &  $-44.02$      & 12.48	& \nodata      	&  381	&  2000-05-29 &    A1180101  &	      \\
	       		& L97	         &	    &		     &	0.21  	&		&	&  2001-06-13 &    B0890101  &	      \\
	       		&	         &	    &		     &	        &		&	&	      &    	     &	      \\
\enddata
\tablenotetext{a}{AV\,18 and AV\,488 are  SMC stars, all the others are LMC stars.}
\tablenotetext{b}{$\Delta t$ between successive observations.}
\tablenotetext{c}{Radial velocity variation detected or suspected.}
\tablerefs{($a$) Sources of spectral type: 
C86$=$\citet{conti86};
F88, F91$=$\citet{fitz88,fitz91};
J01$=$\citet{jaxon01};
L97$=$\citet{lennon97};
W77, W95, W02$=$\citet{walborn77,walborn95,walborn02};
Wpc$=$Walborn (private communication, 2001).
$v_\infty$ from ($b$) \citet{prinja98}, ($c$) \citet{massa02}, 
($d$) \citet{patriarchi}, see also \citet[1800 \km]{bianchi00} for Sk\,$-$67\degr111.
}
\end{deluxetable}

\newpage
\begin{deluxetable}{lllll}
\tablecolumns{5}
\tablewidth{0pc} 
\tabletypesize{\scriptsize}
\tablecaption{Characteristics of Wind Profile Variability in  \ion{O}{6}, \ion{S}{4}, and \ion{P}{5} \tablenotemark{a} \label{t3}} 
\tablehead{\colhead{Star}    &   \colhead{Sp. Type}&   \colhead{\ion{O}{6}}&   \colhead{\ion{S}{4}}&   \colhead{\ion{P}{5}}}
\startdata
\cutinhead{MW sample}
HDE\,303308     	&  O3\,V((f))   & v. $[1100,2700]$	& n.v.	   	&	n.v.     	\\
HD\,64568       	& O3\,V((f$^*$))& n.v.	 		& n.v.	   	&	n.v.     	\\
CPD-59\degr2600		&  O6\,V((f))   & v.(n)	 		& n.v.	   	&	n.v.     	\\
CPD-59\degr2603		&  O7\,V        & v. $[1500,2350]$	& n.v.	   	&	n.v.     	\\
HD\,152590      	&  O7.5\,V      & v.	$[1100,1900]$ 	& n.v.	   	&	n.v.     	\\
HD\,61347       	&  O9\,Ib       & v.	$[900,1900]$ 	& v. $[200,1600]$&	n.v. (n)    	\\
HD\,210809      	&  O9\,Iab      & v.	$[700,2200]$ 	& v. $[300,2200]$&	v. $[300,2200]$	\\
HD\,153426      	&  O9\,II       & n.v.(n)	 	& n.v.  	&	n.v.     	\\
HD\,91651       	&  O9\,V:n      & n.v.	 		& n.v.	   	&	n.v.     	\\
HD\,92554       	&  O9.5\,IIn    & v.	 $[800,1400]$	& v. $[700,1300]$&	n.v. (n)    	\\
HD\,168941      	&  O9.5\,II-III & v.	 $[1000,2100]$	& n.v.	   	&	n.v.     	\\
HD\,156292      	&  O9.5\,III    & n.v.	 		& n.v.	   	&	n.v.     	\\
HD\,163892      	&  O9\,IV ((n)) & v.	 $[950,1550]$ 	& n.v.	  	&	n.v. 	    	\\
HDE\,308813     	&  O9.5\,V      & n.v.	 		& n.v.	   	&	n.v.     	\\
HD\,186994      	&   B0.2IV      & n.v.	 		& n.v.	   	&	n.v.     	\\
CPD$-$72\degr1184	& B0\,III       & n.v.	 		& n.v.	   	&	n.v.     	\\
HD\,192035		&  B0\,III-IV(n)& v.	 $[800,1050]$  	& n.v.	  	&	n.v. 	    	\\
HD\,47417		&  B0\,IV       & v.	 $[650,1025]$ 	& n.v.	  	&	n.v. 	    	\\
HD\,203374A     	&  B0\,IVpe     & n.v.	 		& n.v.	   	&	n.v.     	\\
BD\,+53\degr2820	&  B0\,IVn      & n.v.	 		& n.v.	   	&	n.v.     	\\
HD\,191495		&  B0\,IV-V     & v.(n)	 		& n.v.	   	&	n.v.     	\\
HD\,195965		&  B0\,V        & n.v.	 		& n.v.	   	&	n.v.     	\\
HD\,207538      	&  B0.2\,V      & n.v.	 		& n.v.	   	&	n.v.     	\\
HD\,187459		&  B0.5\,Ib     & v. $[750,1550]$	& v. $[750,1550]$ &	v. $[750,1550]$	\\
HD\,224151		&  B0.5\,II     & v. $[400,1100]$	& n.v.	  	&	n.v. 	    	\\
HDE\,332407		&  B0.5\,III:   & v.(n)	 		& n.v.	   	&	n.v.     	\\
HD\,161807		&  B0.5\,III:n  & v. $[975,1200]$ 	& n.v.	  	&	n.v. 	    	\\
HD\,172140		&  B0.5\,III    & v. $[1000,1400]$  	& n.v.	  	&	n.v. 	    	\\
HD\,62866		&  B0.5\,IIIn   & v. $[800,1150]$	& n.v.	  	&	n.v. 	    	\\
HD\,97913		&  B0.5\,IVn    & v. $[1200,1650]$ 	& n.v.	  	&	n.v. 	    	\\
BD\,+35\degr4258	&  B0.5\,Vn     & v. $[600,950]$ 	& n.v.	  	&	n.v. 	    	\\
HD\,148422		&  B1\,Ia       & v.(n)		 	& n.v.  	&	n.v.     	\\
HD\,191877		&  B1\,Ib       & v.(n)		 	& n.v.  	&	n.v.     	\\
HDE\,235783		&  B1\,Ib       & v.(n)			& n.v. 	 	&	 n.v.		\\
BD\,+48\degr3437	&  B1\,Iab      & n.v.	 		& n.v.	   	&	n.v.     	\\
HD\,225757		&  B1\,IIIn     & n.v.	 		& n.v.	   	&	n.v.     	\\
HD\,91597		&  B1\,IIIne    & v. $[900,1900]$ 	& n.v.	  	&	n.v. 	    	\\
BD\,+52\degr3210	&  B1\,V    	& n.v.	 		& n.v.	   	&	n.v.     	\\
HD\,73   		&  B1.5\,IV     & n.v.(n) 		& n.v.	   	&	n.v.     	\\
HD\,202347		&  B1.5\,V      & n.v.	 		& n.v.	   	&	n.v.     	\\
BD\,+53\degr2885	&  B2\,III      & n.v.	 		& n.v.	   	&	n.v.     	\\
HD\,47961       	&  B2\,V        & n.v.	 		& n.v.	   	&	n.v.     	\\
HD\,51013       	&  B3\,V        & n.v.	 		& n.v.	   	&	n.v.     	\\
HD\,47777       	&  B3\,V        & n.v.	 		& n.v.	   	&	n.v.     	\\
\cutinhead{MC sample}
Sk\,$-$67\degr167	&  O4\,Inf$^+$  & n.v.	 		& n.v.	   	&	n.v.     	\\
Sk\,$-$71\degr45	&  O4-5\,IIIf   & v.(n)			& n.v. 	 	&	 n.v.		\\
Sk\,$-$67\degr111	&  O6:\,Iafpe   & v.			& v. $[500,1600]$&	 v. $[500,1600]$		\\
BI\,208			&  O7\,V$z$     & v. $[800,1900]$	& n.v.	  	&	n.v. 	    	\\
BI\,272		        &  O7:\,III-II: & v. $[400,2600]$	& n.v.	  	&	n.v. 	    	\\
Sk\,$-$67\degr101	&  O8\,II((f))  & v. $[1300,1525]$	& n.v.	  	&	n.v. 	    	\\
BI\,173		        &  O8\,II:      & v.	 		& n.v.	  	&	n.v. 	    	\\
Sk\,$-$67\degr191	&  O8\,V        & v. $[1150,1900]$  	& n.v.	  	&	n.v. 	    	\\
Sk\,$-$68\degr03	&  O9\,I        & n.v.	 		& n.v.	   	&	n.v.     	\\
Sk\,$-$69\degr124	&  O9\,Ib:      & v. $[600,1600]$  	& v.(n)	  	&	n.v. 	    	\\
Sk\,$-$65\degr44	&  O9\,V        & v. $[1200,1600]$	& n.v.	  	&	n.v. 	    	\\
Sk\,$-$67\degr05	&  O9.7\,Ib     & v.	 		& v.	  	&	n.v. 	    	\\
Sk\,$-$65\degr21	&  O9.7\,Iab    & v.	 		& v.	  	&	n.v. 	    	\\
Sk\,$-$71\degr08	&  O9\,II       & n.v.	 		& n.v.	   	&	n.v.     	\\
Sk\,$-$68\degr41	&  B0.5\,Ia     & v. $[100,1200]$ 	& v. $[100,1200]$&	n.v. 	    	\\
AV\,488			&  B0.5\,Iaw    & n.v.	 		& n.v.	   	&	n.v.     	\\
Sk\,$-$67\degr28	&  B0.7\,Ia     & n.v.	 		& n.v.	   	&	n.v.     	\\
Sk\,$-$70\degr85	&  B0\,I        & v.	 		& v. $[100,1200]$&	n.v. 	    	\\
Sk\,$-$68\degr75	&  B1\,I        & n.v.	 		& n.v.	   	&	n.v.     	\\
Sk\,$-$70\degr120	&  B1\,Ia       & n.v.	 		& n.v.	   	&	n.v.     	\\
Sk\,$-$67\degr14	&  B1.5\,Ia     & n.v.	 		& n.v.	   	&	n.v.     	\\
AV\,18			&  B2\,Ia       & n.v.	 		& n.v.	   	&	n.v.     	\\
\enddata
\tablenotetext{a}{v.: wind variable; n.v.: wind not variable; (n)$=$noisy. The approximate 
range of the profile variation is given in square brackets. The values are expressed
in \km\ and should be multiplied by $-1$.
For the Magellanic stars, this range can be more uncertain because of possible overlapping of the \ion{O}{6} 
profiles; and in cases where it is completly uncertain, we do not indicate any values.}
\end{deluxetable}

\newpage

\begin{deluxetable}{llccc}
\tablecolumns{5}
\tablewidth{0pc} 
\tabletypesize{\scriptsize}
\tablecaption{\ion{O}{6} Wind Absorption \label{t4}} 
\tablehead{\colhead{Star}    &   \colhead{Sp. Type}&   \colhead{$v_\infty$}&   \colhead{$v_{\rm abs}$}&   \colhead{Strength} \\
\colhead{}    &   \colhead{}&   \colhead{(\km)}&   \colhead{(\km)}&   \colhead{}}
\startdata
\cutinhead{MW sample}
HD\,152590      &  O7.5\,V     & 1785	 &    1700$^a$ 	& strong	\\
	        &  	       & 	 &    1300  	& weak		\\
HD\,61347       &  O9\,Ib      & 1775	 &    1250	& weak		\\
HD\,153426      &  O9\,II      & 2200	 &    2150 	& strong	\\
HD\,92554       &  O9.5\,IIn   & 1260	 &    1125 	& strong	\\
HD\,168941      &  O9.5\,II-III& 1795	 &    1625$^a$ 	& strong	\\
	        &  	       & 	 &    1250  	& weak		\\
HD\,156292      &  O9.5\,III   & 1355	 &    1300 	& strong	\\
	        &  	       & 	 &    1100  	& weak		\\
HD\,163892      &  O9\,IV ((n))& 1405    &    1320 	& medium	\\   
	        &  	       & 	 &    1175  	& medium	\\
CPD$-$72\degr1184& B0\,III     &\nodata  &    1075 	& medium	\\
BD\,+53\degr2820&  B0\,IVn     &\nodata  &    1200 	& medium	\\
HD\,187459	&  B0.5\,Ib    &1185	 &    1125 	& strong	\\
HD\,224151	&  B0.5\,II    &1280	 &    1250 	& strong	\\
HDE\,332407	&  B0.5\,III:  &\nodata  &    1275 	& medium	\\
HD\,161807	&  B0.5\,III:n &\nodata  &    1120 	& weak		\\
HD\,62866	&  B0.5\,IIIn  &\nodata  &    980 	& weak		\\  
 	        &  	       & 	 &    1130  	& weak		\\
HD\,191877	&  B1\,Ib      &\nodata  &    1000 	& strong	\\  
HD\,91597	&  B1\,IIIne   &\nodata  &    1170 	& weak		\\  
 	        &  	       & 	 &    1370  	& weak		\\
\cutinhead{MC sample}
Sk\,$-$67\degr101&  O8\,II((f))&2005     &    950 	& weak		\\   
Sk\,$-$67\degr191&  O8\,V      &1750     &   1700$^a$	& strong	\\   
Sk\,$-$69\degr124&  O9\,Ib:    &1430     &   1150 	& medium	\\   
	         &  	       & 	 &   1350  	& medium	\\
Sk\,$-$67\degr05 & O9.7\,Ib    &1665     &   1000 	& strong	\\   
Sk\,$-$65\degr21&  O9.7\,Iab   &1330     &   1000 	& medium	\\   
Sk\,$-$70\degr85&  B0\,I       &\nodata  &   850 	& strong	\\   
Sk\,$-$68\degr41&  B0.5\,Ia    &865      &   800 	& strong	\\   
\enddata
\tablecomments{$v_{\rm abs}$ corresponds to the deepest part of the \ion{O}{6} wind absorption feature.
The strength corresponds to the strength of the \ion{O}{6} $\lambda$1038 wind absorption feature,
see \S~\ref{dac} for more details. 
($a$) The \ion{O}{6} $\lambda$1038 wind absorption feature is blended with the interstellar \ion{O}{6} $\lambda$1032,
but the latter is too wide and strong to be entirely interstellar.
}
\end{deluxetable}

\end{document}